\documentclass{aastex}
\newif\ifgrayscale
\grayscaletrue % to compile with grayscale images
\newif\ifincludeplots
\includeplotstrue %uncomment to compile with figures. you also need to comment out the line below.
\newcommand{\sameyear}[1]{}

\newcommand{\eg}{{\it e.g.}}

\newcommand{\ie}{{\it i.e.}}

\newcommand{\etal}{{\it et~al.}}

\newcommand{\x}{ \; \mathrm{x} \;}

\newcommand{\mps}{\,\mathrm{m\;s^{-1}}}

\newcommand{\gpcmc}{\,\mathrm{g \; cm^{-3}}}

\newcommand{\au}{\,\mathrm{au}}

\newcommand{\myr}{\,\mathrm{Myr}}

\newcommand{\pv}{p_V}

\newcommand{\tdeg}{^{\circ}}

\newcommand{\FY}{{1993~FY$_{12}\;$}}
\newcommand{\FYnospace}{{1993~FY$_{12}$}}
\usepackage{epstopdf}
\usepackage{breqn}

\begin{document}

%\title{Modification of asteroid family Yarkovsky V-shapes due to the dependence of thermal inertia on asteroid size}
\title{Initial velocity V-shapes of young asteroid families}

\author{Bryce T. Bolin\altaffilmark{1,2} \footnote{bryce.bolin@oca.eu},
Kevin J. Walsh\altaffilmark{3},
Alessandro Morbidelli\altaffilmark{1}, Marco Delb\'{o}\altaffilmark{1}}

\received{---}
\accepted{---}

\slugcomment{68 Pages, 25 Figures, 2 Tables}
\altaffiltext{1}{Laboratoire Lagrange, Universit\'e C\^ote d'Azur, Observatoire de la C\^ote
d'Azur, CNRS, Blvd. de l'Observatoire, CS 34229, 06304 Nice cedex 4,
France}
\altaffiltext{2}{B612 Asteroid Institute,  20 Sunnyside Ave,  Suite 427, Mill
Valley, CA, 94941, United States}
\altaffiltext{3}{Southwest Research Institute, 1050 Walnut St. Suite 300, Boulder, CO 80302, United States}

\shorttitle{Asteroid family initial velocity V-shapes}
\shortauthors{Bolin \etal}

% ABSTRACT -----------------------------------------------

\begin{abstract}
Ejection velocity fields of asteroid families are largely unconstrained due to the fact that members disperse relatively quickly on Myr time-scales by secular resonances and the Yarkovsky effect. The spreading of fragments in $a$ by the Yarkovsky effect is indistinguishable from the spreading caused by the initial ejection of fragments. By examining families $<$20 Myrs-old, we can use the V-shape identification technique to separate family shapes that are due to the initial ejection velocity field and those that are due to the Yarkovsky effect. $<$20 Myr-old asteroid families provide an opportunity to study the velocity field of family fragments before they become too dispersed. Only the Karin family's initial velocity field has been determined and scales inversely with diameter, $D^{-1}$. We have applied the V-shape identification technique to constrain young families' initial ejection velocity fields by measuring the curvature of their fragments' V-shape correlation in semi-major axis, $a$, vs. $D^{-1}$ space. Curvature from a straight line implies a deviation from a scaling of $D^{-1}$. We measure the V-shape curvature of 11 young asteroid families including the \FYnospace, Aeolia, Brangane, Brasilia, Clarissa, Iannini, Karin, Konig, Koronis(2), Theobalda and Veritas asteroid families. We find that the majority of asteroid families have initial ejection velocity fields consistent with $\sim D^{-1}$ supporting laboratory impact experiments and computer simulations of disrupting asteroid parent bodies.
\end{abstract}

{\it Key words}: minor planets, asteroids, general

\maketitle

% RUNNING HEAD & CORRESPONDENCE -----------------------------------------------

\newpage
%\begin{center}
\noindent {\bf Proposed Running Head:} Initial velocity V-shapes of young Main Belt asteroid families \\
%\end{center}

\vspace{10cm}

\noindent {\bf Editorial correspondence to:} \\
Bryce Bolin \\
Observatoire de la Cote d'Azur \\
Boulevard de l'Observatoire \\
CS 34229 \\
06304 Nice, France \\
Phone: +33 04 92 00 30 81 \\
Fax: +33 (0) 4 92 00 30 33 \\
E-mail: bbolin@oca.eu

% INTRODUCTION -----------------------------------------------------------

\section{Introduction}
\label{s.Introduction}

Asteroid families are formed as a result of collisional disruptions and cratering events on larger parent bodies \citep[\eg ][]{ Durda2004, Michel2015}. Although dispersed in space, the family members typically cluster in their proper orbital elements, semi-major axis $(a)$, eccentricity $(e)$ and inclination $(i)$, close to that of the parent body \citep[\eg ][]{Hirayama1918, Zappala1995,Nesvorny2015a} and share similar spectral and reflectance properties \citep[][]{Cellino2002a, Masiero2013, deLeon2016}.

It was thought that asteroid fragments remained stationary in orbital elements space after their disruption \citep[][]{Zappala1996, Cellino1999} requiring large ejection velocities to explain the wide dispersion of asteroid family fragments in orbital elements space. Impact simulations and asteroids observed on temporary, unstable orbits as well as family fragments leaking through mean motion and secular resonances provided evidence that asteroids' orbits were modified due to  recoil from anisotropic surface emission of thermal photons, \ie, the Yarkovsky effect was responsible for the large dispersion of asteroids in orbital elements space \citep[][]{Michel2001,Bottke2001}. There is a degeneracy between the contribution of the initial spreading of an asteroid family fragments' orbital elements caused by the initial ejection of fragments and the contribution caused by the subsequent drift in $a$ caused by the Yarkovsky effect. This degeneracy can only be broken in special cases, such as asteroids leaking through resonances or families that are too disperse to show the imprint of the initial ejection of fragments. 

In addition to the cases above, young asteroid families can provide an opportunity to determine how ejection velocities of asteroid family fragments are distributed as a function of their size. The ejection velocities of Karin and Koronis asteroid family fragments have been measured to scale inversely with $D$ by the use of Gauss' equations \citep[][]{Zappala1990} or by linking the $i$ distribution of family fragments to their out of plane velocities \cite[][]{Nesvorny2002b, Carruba2016e}. In this paper we study the least dispersed families in the asteroid belt, which most likely are the youngest, and examine how the spreading of their members depends on size. In an accompanying paper \citep[][]{Bolin2017b} we show that semi-major axis spreading of the oldest and most disperse families follows a different dependence with the fragments' size. Thus, we demonstrate that the shape of a family in the $a$ vs. $\frac{1}{D}$ space is a generic way to break the degeneracy between initial ejection velocity and Yarkovsky evolution and can be used to tell the relative contribution of each of these processes. 

\section{Initial velocity field V-shapes}
\label{s.Vshapedefinition}
The displacement in $\Delta a \; = \; \left | a - a_c \right |$ after the disruption of a parent body, where $a_c$ is the location in $a$ of the parent body, is a function of $V_T$, the transverse velocity component of the ejected fragment and its parent body's mean motion, $n$   \citep[][]{Zappala1996}
\begin{equation}
\label{eqn.deltaaejectionV_Tvs_a_c}
\left | a - a_c \right | \; = \; \frac{2}{n}\; V_{T} 
\end{equation}
 The initial ejection of the family fragments should result in a symmetric V-shaped spread of fragments in $a$ vs. the reciprocal of the diameter, $\frac{1}{D} \; = \; D_r$, space because $V_T$ scales inversely with asteroid diameter $D$ \citep[][]{Cellino1999,Vokrouhlicky2006b}
\begin{equation}
\label{eqn.deltaaejectionD_r_theta_vs_a_c}
\left | a - a_c \right | \; = \; \frac{2}{n}\;V_{EV} \; \left ( \frac{D_0}{D} \right )^{\alpha_{EV}} \; \mathrm{cos}(\theta)
\end{equation}
$D_0$ is equal to 1329 km where $V_{EV}$ is a parameter that describes the width of the fragment ejection velocity distribution \citep[][]{Michel2004, Nesvorny2006a,  Vokrouhlicky2006b,Vokrouhlicky2006a, Durda2007}. $V_{EV}$ for known asteroid families such as Karin and Erigone range between 15 and 50 $\mps$ \citep[][]{Nesvorny2006a, Bottke2007, Carruba2011, Masiero2012b, Nesvorny2015a}. 

$V_{EV}$ is determined by modedlling the initial ejection of fragments according to Eq.\ref{eqn.deltaaejectionD_r_theta_vs_a_c} where $\alpha_{EV}$ is the exponent scaling $V_{EV}$ with $D$ \citep[][]{Vokrouhlicky2006b, Vokrouhlicky2006a}. $\alpha_{EV}$ = 1 would imply a simple $\frac{1}{D}$ $V_{EV}$ dependence. Modedlling of the ejection of fragments is done for asteroid families younger than $\lesssim$20 Myrs where the Yarkovsky effect has not had enough time to modify the $a$ of the fragments such as for the Karin asteroid family \citep[][]{Nesvorny2006a}. The modedlling of the initial ejection velocities of the fragments includes evolution of family fragments' $a$ according to the Yarkovsky and YORP effects \citep[][]{Bottke2006, Vokrouhlicky2015}. Additional constraints to the initial ejection velocity field can be provided by the $i$ distribution of an asteroid family, which is supposed to remain essentially unaltered during the Yarkosky evolution \citep[][]{Carruba2016d,Carruba2016e}. For cases where $V_{EV}$ can not be determined, the escape velocity of the asteroid family parent body is used for $V_{EV}$ \citep[][]{Vokrouhlicky2006b,Walsh2013}. The escape velocity of the parent body is a good estimate for $V_{EV}$ because most particles are ejected from their parent bodies at velocities around the escape velocity of the parent body in numerical impact simulations \citep[][]{Durda2007,Sevecek2017}.

 Karin and Koronis asteroid family data suggest that $\alpha_{EV} \simeq 1.0 $ \citep[][]{Nesvorny2002b, Nesvorny2006a, Carruba2016e} while analytical calculations of $\alpha_{EV}$ can be as high as $\simeq$ 1.5 \citep{Cellino1999}. $\mathrm{cos}(\theta)$ is the angle of the fragment's velocity relative the transverse direction of the parent body's orbit. In Eq.~\ref{eqn.deltaaejectionD_r_theta_vs_a_c}, $\mathrm{cos}(\theta)$ is expected to be uniformly distributed between -1 and 1. We assume that the number of fragments is high enough that the V-shape's edge is defined by fragments with $\mathrm{cos}(\theta) \; \simeq \; 1.0$ or -1.0 in Eq.~\ref{eqn.deltaaejectionD_r_theta_vs_a_c}. The $V_{EV}$ of fragments interior to the V-shape border with the same value of $\left | \mathrm{cos}(\theta) \right | \; < \; $ 1 will scale with $D$ by the same $\alpha_{EV}$ as fragments on the V-shape border. The left side of Eq.~\ref{eqn.deltaaejectionD_r_theta_vs_a_c} is decreased for V-shapes with fewer fragments by a factor of $\frac{2}{\pi}$ because $\frac{2}{\pi}$ is the average value of $\mathrm{cos}(\theta)$ between the intervals 0 and $\frac{\pi}{2}$ and $\frac{\pi}{2}$ and $\pi$ resulting in a distorted value of $\alpha$. 

We re-write Eq.~\ref{eqn.deltaaejectionD_r_theta_vs_a_c} in $a$ vs. $D_r$ space with $\mathrm{cos}(\theta) \; = \; 1.0$ to obtain $D_r$ as a function of $a,\;a_c\;,n,\;V_{EV}$ and $\alpha_{EV}$
\begin{equation}
\label{eqn.deltaaejection_a_vs_D_r}
D_r(a,a_c,n,V_{EV},\alpha_{EV}) \; = \; \frac{1}{D_0} \; \left ( \frac{\left | a - a_c \right | \; n}{2 \; V_{EV}} \right )^{\frac{1}{\alpha_{EV}} }
\end{equation}

The spread in family fragments by their initial ejection from the parent body has the same functional form of the spreading of family fragments caused by the Yarkovksy effect \citep[][]{Bottke2006, Vokrouhlicky2015}. 
\begin{equation}
\label{eqn.apvDvsCmod}
D_r(a,a_c,C_{\alpha},\pv,\alpha) \; = \frac{\sqrt{\pv}}{D_0}\; \left(\frac{\left | a - a_c \right |}{ C_{\alpha}}\right)^{\frac{1}{\alpha}}
\end{equation}
$C_\alpha$ is the width of the V-shape for a specific value of $\alpha$. $C_{\alpha}$ is normalised to the width of the V-shape with $\alpha \; = \; 1.0$\begin{equation}
\label{Cnorm}
C = C_{\alpha}\left( \frac{\sqrt{\pv}}{1329} \right )^{1-\alpha}
\end{equation}
Eq.~\ref{eqn.apvDvsCmod} is re-written in terms of $(a,a_c,C,\pv,\alpha)$ by using Eq.~\ref{Cnorm}
\begin{equation}
\label{eqn.apvDvsCfinal}
D_r(a,a_c,C,\pv,\alpha) \; = \; \left(\frac{\left | a - a_c \right | \sqrt{\pv}}{ D_0 \; C}\right)^{\frac{1}{\alpha}}
\end{equation}
The parameters, $\alpha_{EV}$ and $\alpha$ in Eqs.~\ref{eqn.deltaaejectionD_r_theta_vs_a_c} and \ref{eqn.apvDvsCfinal}, describing the shape of family V-shapes in $a$ vs. $D_r$ caused by the $D$ dependent initial ejection of fragments or Yakovsky force are functionally equivalent.

We combine Eqs.~\ref{eqn.deltaaejection_a_vs_D_r} and \ref{eqn.apvDvsCfinal} to obtain the spread of the V-shape, C, caused by the initial ejection of fragments as a function of $n, \; V_{EV}, \;  \pv,$ and $\alpha_{EV}$  
\begin{equation}
\label{eqn.VEVvsCalphaFinal}
C(n, V_{EV}, \pv, \alpha_{EV}) \; = \; \frac{2 \cdot D_0^{\alpha_{EV} -1} }{n} \; V_{EV} \; \sqrt{\pv}
\end{equation}

We apply the techniques to identify asteroid family Yarkovsky V-shapes as defined in \citet[][]{Bolin2017} to measure the $\alpha$ of an asteroid family's initial ejection velocity V-shape because the shape of the initial ejection velocity field as described by Eq.~\ref{eqn.deltaaejection_a_vs_D_r} is similar to shape acquired by the Yarkovsky effect described by Eq.~\ref{eqn.apvDvsCfinal}. 

\subsection{V-shape identification technique and measurement of $\alpha$} 
\label{s.v-shapeidentificationandalpha}
The identification of the family V-shape is performed  by determining $a_c$, $C$ and $\alpha$ for a family V-shape according to Eq.~\ref{eqn.apvDvsCfinal} in $a$ vs. $D_r$ space using a modified version of the border method from \citet[][]{Bolin2017}.
\begin{equation} 
\label{eq.border_method_N_outer}
  N_{out}(a_c,C,dC,\pv,\alpha)
    = \; \frac{\Sigma_j 
       \; w(D_j) \;
       \int\limits_{a_1}^{a_2} da
    	\int\limits_{D_r(a, a_c,C_+, \pv,\alpha)}^{D_r(a, a_c,C, \pv,\alpha)} dD_r
       \; \delta(a_{j}-a)
       \; \delta( D_{r,j}-D_r )}{{\int\limits_{a_1}^{a_2}  da
    	\int\limits_{D_r(a, a_c,C, \pv,\alpha)}^{D_r(a, a_c,C, \pv,\alpha)} \; dD_r}}
\end{equation}
\begin{equation} 
\label{eq.border_method_N_inner}
  N_{in}(a_c,C,dC,\pv,\alpha)
    = \; \frac{\Sigma_j 
       \; w(D_j) \;
       \int\limits_{a_1}^{a_2} da
    	\int\limits_{D_r(a, a_c,C, \pv,\alpha)}^{D_r(a, a_c,C_-, \pv,\alpha)} dD_r
       \; \delta(a_{j}-a)
       \; \delta( D_{r,j}-D_r )}{{\int\limits_{a_1}^{a_2}  da
    	\int\limits_{D_r(a, a_c,C, \pv,\alpha)}^{D_r(a, a_c,C, \pv,\alpha)} \; dD_r}}
\end{equation}
Eqs.~\ref{eq.border_method_N_outer} and \ref{eq.border_method_N_inner} are normalised the area in $a$ vs. $D_r$ between the nominal and outer V-shapes defined by $D_r(a, a_c,C_+, \pv,\alpha)$ and $D_r(a, a_c,C, \pv,\alpha)$ in the denominator for Eq.~\ref{eq.border_method_N_outer} and between the nominal and inner V-shapes defined by defined by $D_r(a, a_c,C, \pv,\alpha)$ and $D_r(a, a_c,C_-, \pv,\alpha)$ in the denominator for Eq.~\ref{eq.border_method_N_inner}.

The symbol $\Sigma_j$ in Eqs.\ref{eq.border_method_N_outer} and \ref{eq.border_method_N_inner} indicates summation on the asteroids of the catalogue, with semi-major axis $a_j$ and reciprocal diameter $D_{r,j}$. The symbol $\delta$ indicates Dirac's function, and $a_1$ and $a_2$ are the low and high semi-major axis range in which the asteroid catalogue is considered.  The function $w(D)$ weighs the right-side portions of Eqs.~\ref{eq.border_method_N_outer} and \ref{eq.border_method_N_inner} by their size so that the location of the V-shape in $a$ vs. $D_r$ space will be weighted towards its larger members. The exponent 2.5 is used for $w(D) = D^{2.5}$, in agreement with the cumulative size distribution of collisionally relaxed populations and with the observed distribution for MBAs in the $H$ range $12\; < \;H \;< \;16$ \citep[][]{Jedicke2002}.

\begin{figure}
\centering
\hspace*{-0.7cm}
\ifincludeplots
\includegraphics[scale=0.425]{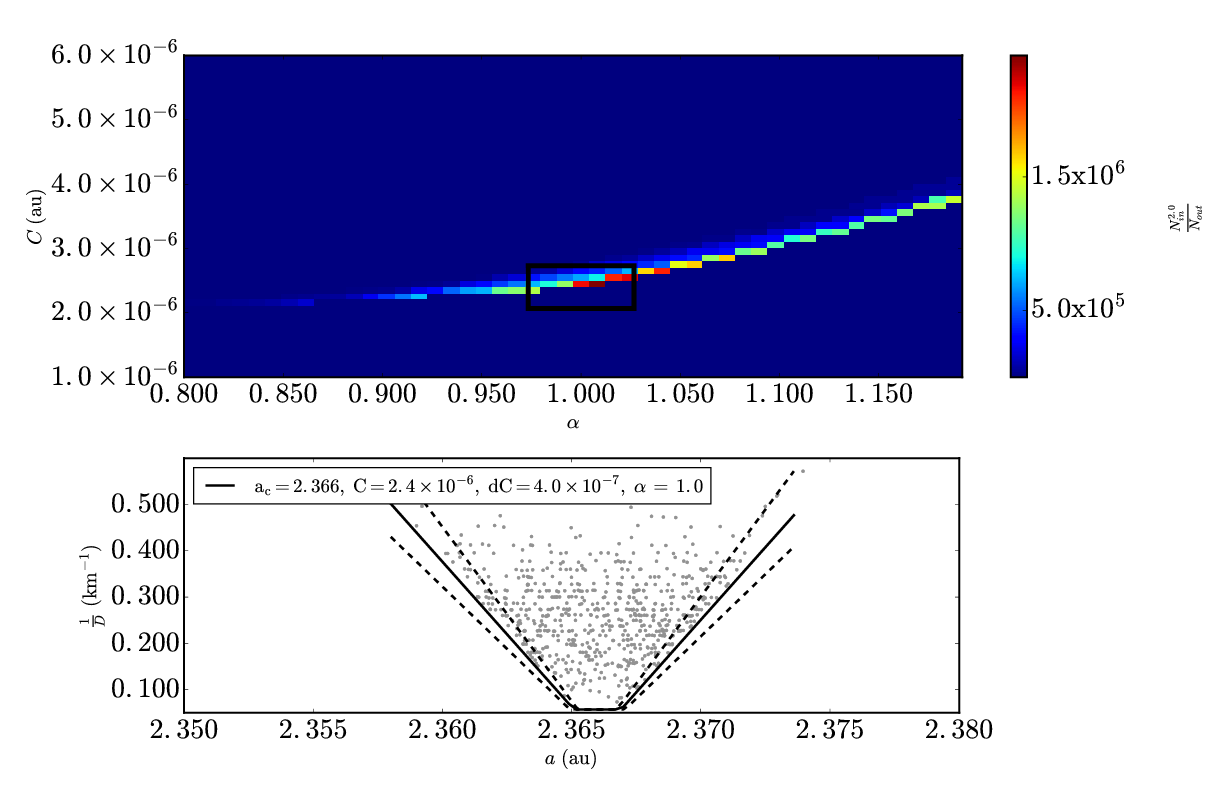}
\else
I am not enabling plots.
\fi
\caption{Application of the V-shape identification to synthetic asteroid family data at Time = 0. (Top panel) The ratio of $N_{out}(a_c,C,dC,\pv,\alpha)^2$ to $N_{in}(a_c,C,dC,\pv,\alpha)$ ratio in the $\alpha$-$C$ range, ($a_c\pm \frac{\Delta \alpha}{2}$,$C\pm \frac{\Delta C}{2}$) where $\Delta \alpha$ is equal to $8.0 \times 10^{-3}$ au and $\Delta C$, not to be confused with $dC$, is equal to $1.0 \times 10^{-9}$ au, for the single synthetic family. The box marks the peak value in $\frac{N_{in}(a_c,C,dC,\pv,\alpha)^2}{N_{out}(a_c,C,dC,\pv,\alpha)}$ for the synthetic family V-shape. (Bottom Panel) $D_r(a,a_c,C,\pv, \alpha)$ is plotted for the peak value with the primary V-shape as a solid line where $\pv = 0.05$. The dashed lines mark the boundaries for the area in $a$ vs. $D_r$ space for $N_{in}$ and $N_{out}$ using Eq.~\ref{eqn.apvDvsCfinal},  $D_r(a,a_c,C\pm dC,\pv,\alpha)$ where $a_c$ = 2.366 au and $dC \; = \; 4.0 \x 10^{-7}$ au.}
\label{fig.synErig0Myrs}
\end{figure}

\citet{Walsh2013} found that the borders of the V-shapes of the Eulalia and new Polana  family could be identified by the peak in the ratio $\frac{N_{in}}{N_{out}}$ where $N_{in}$ and  ${N_{out}}$ are the number of asteroids falling between the curves defined by Eq.~\ref{eqn.apvDvsCfinal} assuming $\alpha \; = \; 1.0$ for values $C$ and $C_-$ and  $C$ and $C_+$, respectively, with $C_-=C-dC$ and $C_+=C+dC$. We extend our technique to search for a peak in the ratio $\frac{N_{in}^2}{N_{out}}$, which corresponds to weighting the ratio of $\frac{N_{in}}{N_{out}}$ by the value of $N_{in}$. This approach has been shown to provide sharper results \citep[][]{Delbo2017}. Here we extend the search for a maximum of $\frac{N_{in}^2}{N_{out}}$ to 3 dimensions, in the $a_c$, $C$ and $\alpha$ space. A peak value in $\frac{N_{in}(a_c,C,dC,\pv,\alpha)^2}{N_{out}(a_c,C,dC,\pv,\alpha)}$  as seen in the top panel of Fig.~\ref{fig.synErig0Myrs} for the synthetic astroid family described in Section~\ref{s.synfamily}. For simplicity only the projection on the $\alpha$, $C$ plane compared to \citet[][]{Bolin2017} and \citet[][]{Delbo2017} that used only the projection in the $a_c$, $C$ plane) indicates the best-fitting values of $a_c$, $C$ and $\alpha$ for a family V-shape using Eq.~\ref{eqn.apvDvsCfinal} (bottom panel of Fig.~\ref{fig.synErig0Myrs}).

\begin{figure}
\centering
\hspace*{-0.7cm}
\ifincludeplots
\includegraphics[scale=0.425]{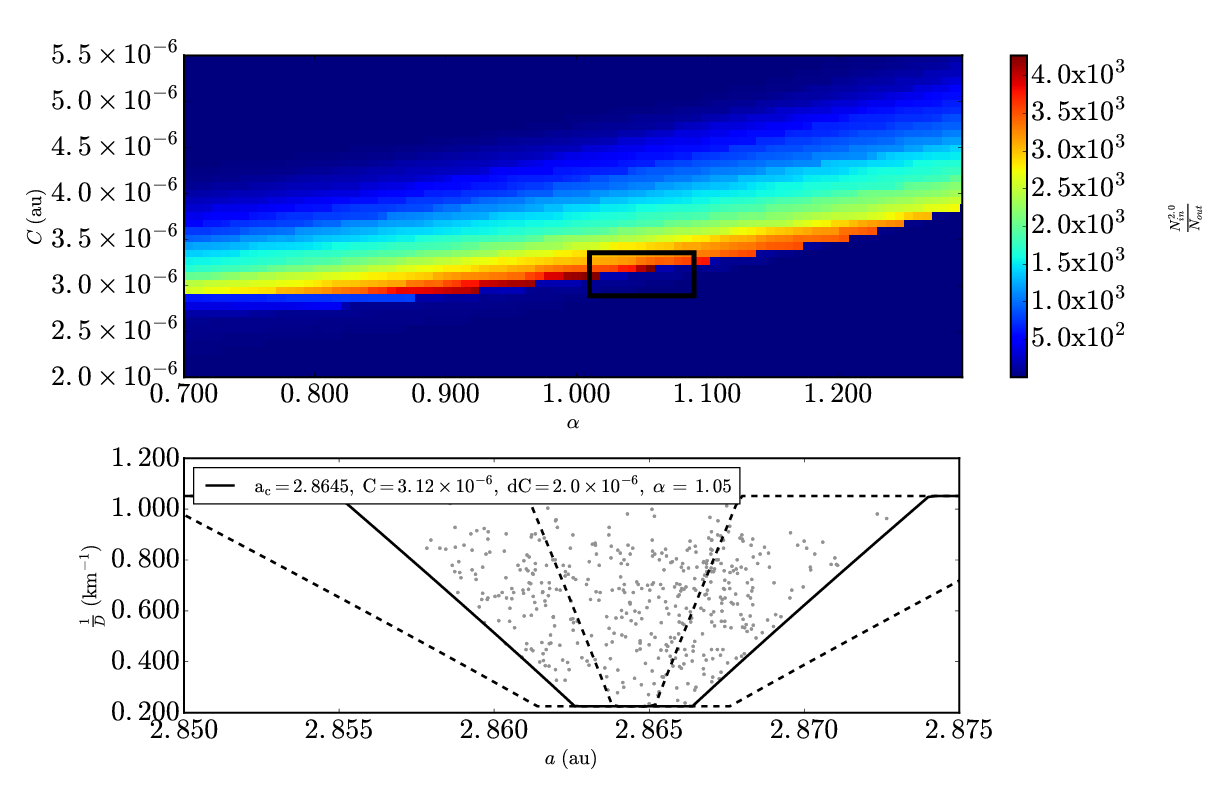}
\else
I am not enabling plots.
\fi
\caption{The same as Fig.~\ref{fig.synErig0Myrs} for Karin asteroid family data from \citet[][]{Nesvorny2015a}. (Top panel) $\Delta \alpha$ is equal to $7.0 \times 10^{-3}$ au and $\Delta C$, is equal to $8.0 \times 10^{-9}$ au. (Bottom Panel) $D_r(a,a_c,C\pm dC,\pv,\alpha)$ is plotted with $\pv = 0.21$, $a_c$ = 2.865 au and $dC \; = \; 2.0 \x 10^{-6}$ au.}
\label{fig.KarinAlph}
\end{figure}

\begin{figure}
\centering
\hspace*{-0.7cm}
\ifincludeplots
\includegraphics[scale=0.425]{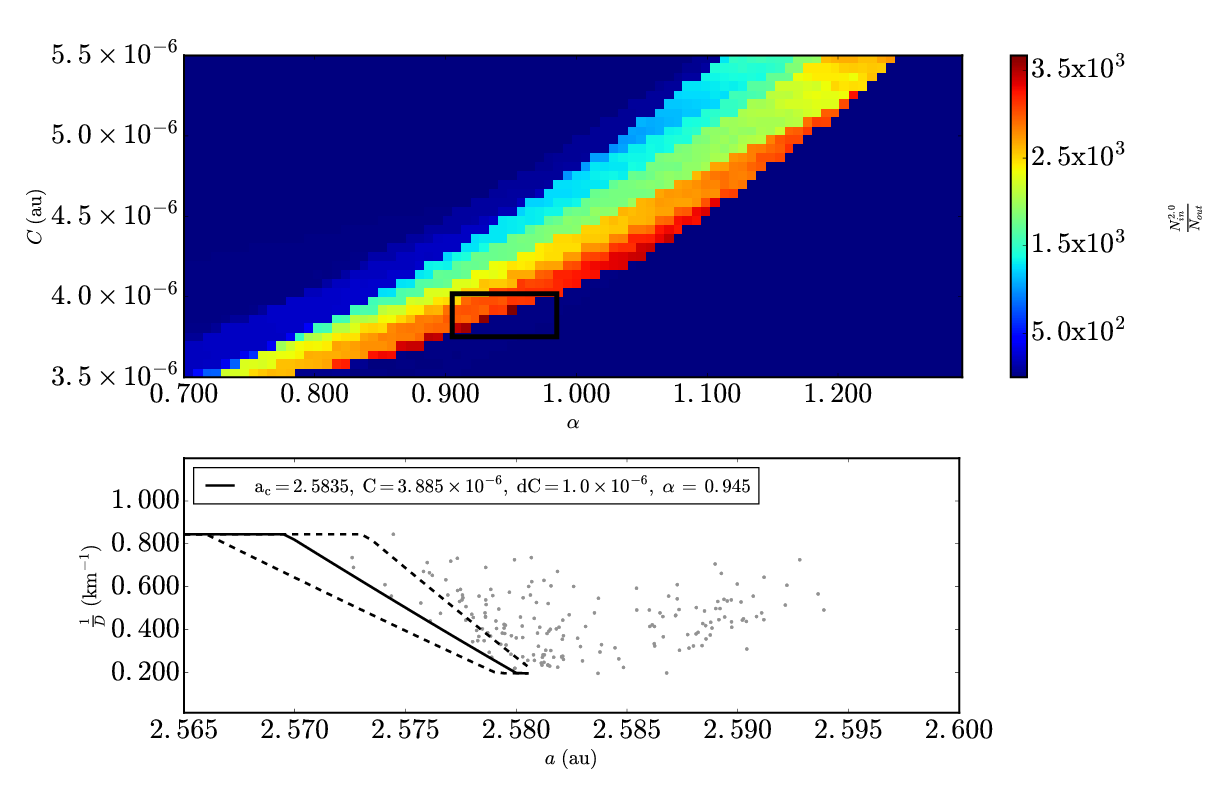}
\else
I am not enabling plots.
\fi
\caption{The same as Fig.~\ref{fig.synErig0Myrs} for Brangane asteroid family data from \citet[][]{Nesvorny2015a}. (Top panel) $\Delta \alpha$ is equal to $7.0 \times 10^{-3}$ au and $\Delta C$, is equal to $5.5 \times 10^{-8}$ au. (Bottom Panel) $D_r(a,a_c,C\pm dC,\pv,\alpha)$ is plotted with $\pv = 0.1$, $a_c$ = 2.584 au and $dC \; = \; 1.0 \x 10^{-6}$ au.}
\label{fig.BranganeAlph}
\end{figure}

The value of $dC$ is used similarly as in \citet[][]{Bolin2017}. The value of dC used depends on the density of asteroids on the family V-shape edge. The value of dC can be a few 10$\%$ of the V-shape's $C$ value if the density of asteroids on the V-shape edge is high such as the case of the Karin family (see the bottom panel of Fig.~\ref{fig.KarinAlph}) and more, up to 40$\sim$50$\%$ if the V-shape edge is more diffuse such as in the case of the Brangane family (see the bottom panel of Figs.~\ref{fig.BranganeAlph}) \citep[][]{Milani2014,Nesvorny2015a,Spoto2015}. The inner and outer V-shapes must be wide enough to include enough asteroids in the inner V-shape and measure a $N_{in}^2$ to $N_{out}$ ratio high enough to identify the family V-shape. The V-shape can include interlopers or asteroids which are not part of the family V-shape if the value used for $dC$ is used is too large \citep[][]{Nesvorny2015a, Radovic2017a}. The V-shape identification technique was tested on families identified by both \citet[][]{Nesvorny2015a} and \citet[][]{Milani2014} to verify that the V-shape $a_c$, $C$ and $\alpha$ determination works on family membership definitions from either database and produces similar results. The V-shape $a_c$, $C$ and $\alpha$ determination technique was tested the \FYnospace, Aeolia, Brangane, Brasilia, Iannini and K{\"o}nig families identified by both \citet[][]{Nesvorny2015a} and \citet[][]{Milani2014} discussed in Sections~\ref{s.FY}, \ref{s.Aeolia}, \ref{s.Brangane}, \ref{s.Brasiliasection}, \ref{s.iannini} and \ref{s.konig}.

\subsection{Contribution to width of young asteroid family V-shapes by the Yarkovsky effect}
\label{s.yarkoage}
The value $C$ obtained by the V-shape $a_c$, $C$ and $\alpha$ determination method in Section~\ref{s.v-shapeidentificationandalpha} includes the contribution of the initial ejection Velocity field from Eq.~\ref{eqn.VEVvsCalphaFinal} and the contribution to $C$ from the Yarkovsky effect \citep[][]{Vokrouhlicky2006b,Nesvorny2015a}
\begin{equation}
\label{eqn.Ccombo}
C \; = \; C_{YE}  \; + \; C_{EV}
\end{equation}
where $C_{YE}$ is the width of the V-shape due to the Yarkovsky effect and $C_{EV}$ is the width of the V-shape due to the inital ejection velocity of fragments. Asteroid family V-shapes with $C \; = \; C_{EV}$ only in Eq.~\ref{eqn.Ccombo} are indistinguishable from asteroid families which have contribution to their value of $C$ from both the Yarkovsky effect and ejection velocity. The nominal value of $C_{EV}$ exceed more than 50$\%$ of $C$ for asteroid families younger than 100 Myrs \citep[][]{Nesvorny2015a,Carruba2016d}. The error on the parent body size and the resultant calculation of $C$ from the parent body's escape velocity can be large enough so that there is a possibility that $C \; - C_{EV} \; \lesssim \; 0$. In this work, we select for analysis all the families for which $C - C_{Ejection\; velocity} \; \lesssim \; 0$, considering that these families are young enough that the contribution of the Yarkovsky effect to the spread in $a$ of the family fragments is minimal and therefore we assume that the value of $C$ obtained with the techniques in Section~\ref{s.v-shapeidentificationandalpha} is equal to $C_{EV}$. 

However, it should be noted that the Yarkovsky effect still affects the displacement in $a$ of even young family members. \citet[][]{Nesvorny2004} and \citet[][]{Carruba2016f} showed that the Yarkovsky drift rate and displacement in $a$ could be determined for Karin family fragments by backwards integrating the orbits of the family fragments backwards in time and measuring the convergence of the ascending node, $\Omega$, and longitude of perihelion, $\varpi$, of the Karin family fragments relative to asteroid Karin while under the influence of the Yarkovsky effect. Although the effect of the Yarkovsky force on the Karin family fragments is strong enough to be detected using the techniques in \citet[][]{Nesvorny2004} and \citet[][]{Carruba2016f}, the displacement in $a$ over the age of the young Karin family is not large enough to affect our assumptions (see Section~\ref{s.karin} about the Karin family as an example).

\subsection{Data set and uncertainties of $\alpha$  measurements}
\subsubsection{Data set}
\begin{figure}
\centering
\hspace*{-0.7cm}
\ifincludeplots
\includegraphics[scale=0.425]{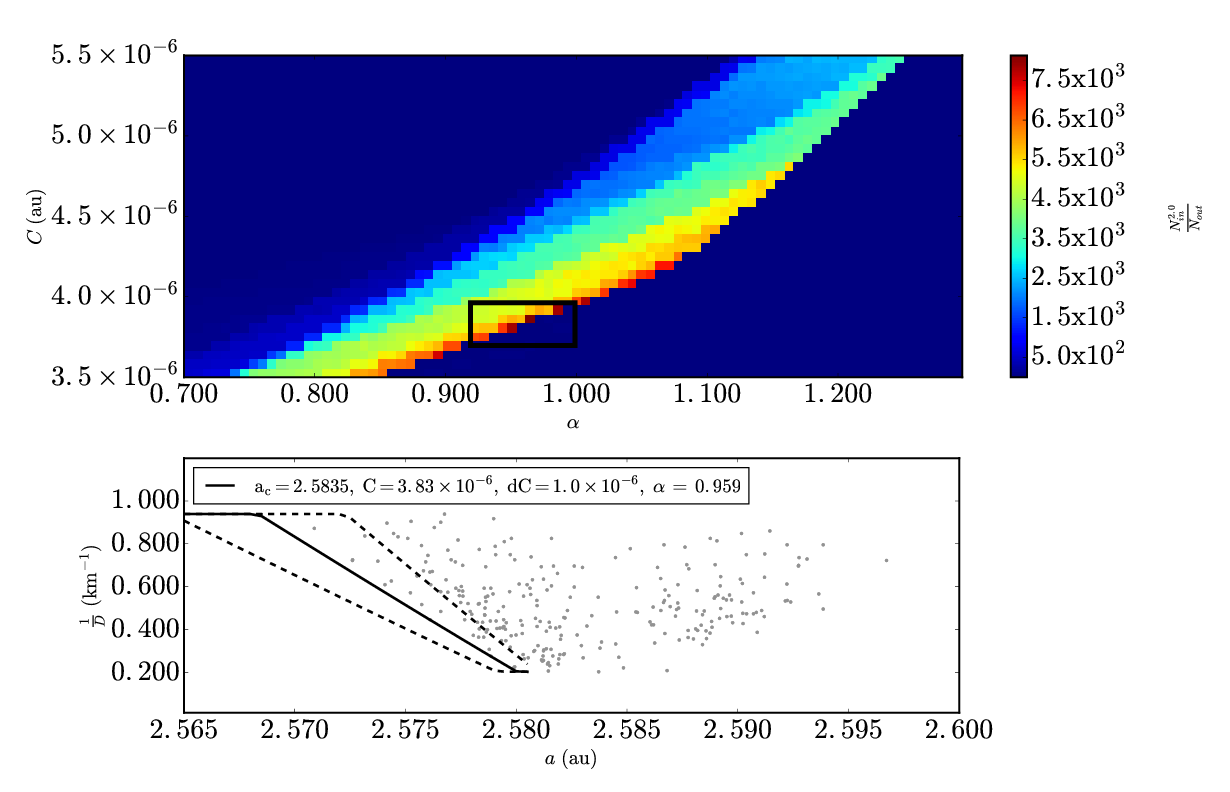}
\else
I am not enabling plots.
\fi
\caption{Same as Fig.~\ref{fig.BranganeAlph}, but repeated for the Brangane family defined by \citet[][]{Milani2014}}
\label{fig.BranganeAlphMilani}
\end{figure}

The data used to measure the V-shapes of asteroid family were taken from the MPC catalogue for the $H$ magnitudes. Family definitions were taken from \citet[][]{Nesvorny2015a}. Asteroid family data for the \FYnospace, Brangane, and Iannini families were used from both \citet[][]{Milani2014} and \citet[][]{Nesvorny2015a} to verify that the V-shape technique provided similar results for $a_c$, $C$ and $\alpha$ for the same family with asteroid data taken from different asteroid family databases. Results from the V-shape technique using asteroid family data from \citet[][]{Nesvorny2015a} were repeated with asteroid family data from \citet[][]{Milani2014} for the \FYnospace, Aeolia, Brangane, Brasilia, Iannini and K{\"o}nig asteroid families as seen for the Brangane family in Figs.~\ref{fig.BranganeAlph} and \ref{fig.BranganeAlphMilani}. Family visual albedo, $\pv$, data from \citet[][]{Masiero2013} and \citet[][]{Spoto2015} were used to calibrate the conversion from $H$ magnitudes to asteroid $D$ using the relation $D = 2.99 \x 10^8 \; \frac{10^{0.2 \; (m_\odot  \; - \; H)}}{\sqrt{\pv}}$ \citep[][]{Bowell1988} where $m_\odot \; = \; -26.76$ \citep[][]{Pravec2007}. Numerically and analytically calculated MBA proper elements were taken from the Asteroid Dynamic Site\footnote{\tt{http://hamilton.dm.unipi.it/astdys/}} \citep[][]{Knezevic2003}. Numerically calculated proper elements were used preferentially and analytical proper elements were used for asteroids, that had numerically calculated elements as of April 2017.

\subsubsection{Uncertainty of $\alpha$}
\label{s.montecarlo}
The value of $\alpha$ located where $\frac{N_{in}(a_c,C,dC,\pv,\alpha)^2}{N_{out}(a_c,C,dC,\pv,\alpha)}$ peaks in $\alpha$ vs. $C$ space represents the best estimate of the $\alpha$ of a asteroid family's V-shape using the nominal $a$ and $D_r$ asteroid values. Different values in the physical properties of asteroids cause a spread in possible $\alpha$ values that measured together are the uncertainty in the measured value of $\alpha$. Changes in asteroids' $D$ is caused by variations in their $H$ magnitude measurements and spread in $\pv$ during the conversion of asteroid $H$ to $D$. In addition to the variety of different possible $D$ values for asteroids and a lack of complete information about the true population of asteroids within a family, the contribution of outliers to a family's $a$ vs. $D_r$ distribution can increase the spread in $\alpha$ values compatible with the family V-shape. We devise the following Monte Carlo procedure to quantify the spread in $\alpha$ measurements of family V-shapes caused by random differences in the asteroid physical properties between family members and incomplete information about the asteroid family member population.

At least 1,200 Monte Carlo trials are completed per family. Some families have significantly more than 1,200 Monte Carlo trials as described in the Appendix if additional CPU time was available. In each trial, the location of the peak $\frac{N_{in}(a_c,C,dC,\pv,\alpha)^2}{N_{out}(a_c,C,dC,\pv,\alpha)}$ value in $\alpha$ vs. $C$ is recorded. Three steps are completed to randomise the asteroid family data from the original $a$ vs. $D_r$ distribution per trial. The first step is to create a resampled data set of family fragments by removing $\sqrt{N}$ objects randomly where $N$ is the number of objects in $a$ vs. $D_r$ space to include variations caused by incomplete knowledge of the asteroid family fragment population. Incompleteness of asteroid family fragments increases for smaller fragments and is more pronounced in the middle and outer portions of the main belt \citep[][]{Jedicke1998,Jedicke2002}. The variation of $\alpha$ caused by the incomplete knowledge of the family fragment population is more weighted towards smaller fragments than larger fragments as a result of the increased incompleteness and greater number of smaller Main Belt asteroids in the asteroid family catalogues

A second step is taken to determine the variation caused by incomplete information in the family fragment population by resampling the fragments' $a$  by their own $a$ distribution per $D_r$ bin. In this step, family fragments are randomised by the semi-major axis distribution of fragments in each $D_r$ bin with a size of 0.001 km$^{-1}$.

The third step is to randomise the measurements of $H$ and $\pv$ of the asteroids by their known uncertainties. Asteroid $H$ values were randomised between 0.2 and 0.3 magnitudes known uncertainties for $H$ values from the Minor Planet Center Catalogue \citep[][]{Oszkiewicz2011, Pravec2012} with an average offset of 0.1 magnitudes consistent for asteroids with $12 \;  < \; H \; < \; 18$ \citep[][]{Pravec2012,Veres2015}. After the $H$ values are randomised, asteroid fragments' $H$ were converted to $D$ using the relation 
\begin{equation}
\label{eq.HtoD}
D = 2.99 \x 10^8 \; \frac{10^{0.2 \; (m_\odot  \; - \; H)}}{\sqrt{\pv}}
 \end{equation}
 from \citet[][]{Harris2002}, and a value of $\pv$ chosen at random for each asteroid using central values and uncertainties per asteroid family from \citet[][]{Masiero2013} and \citet[][]{Spoto2015}.

\begin{figure}
\centering
\hspace*{-1.1cm}
\ifincludeplots
\includegraphics[scale=0.3725]{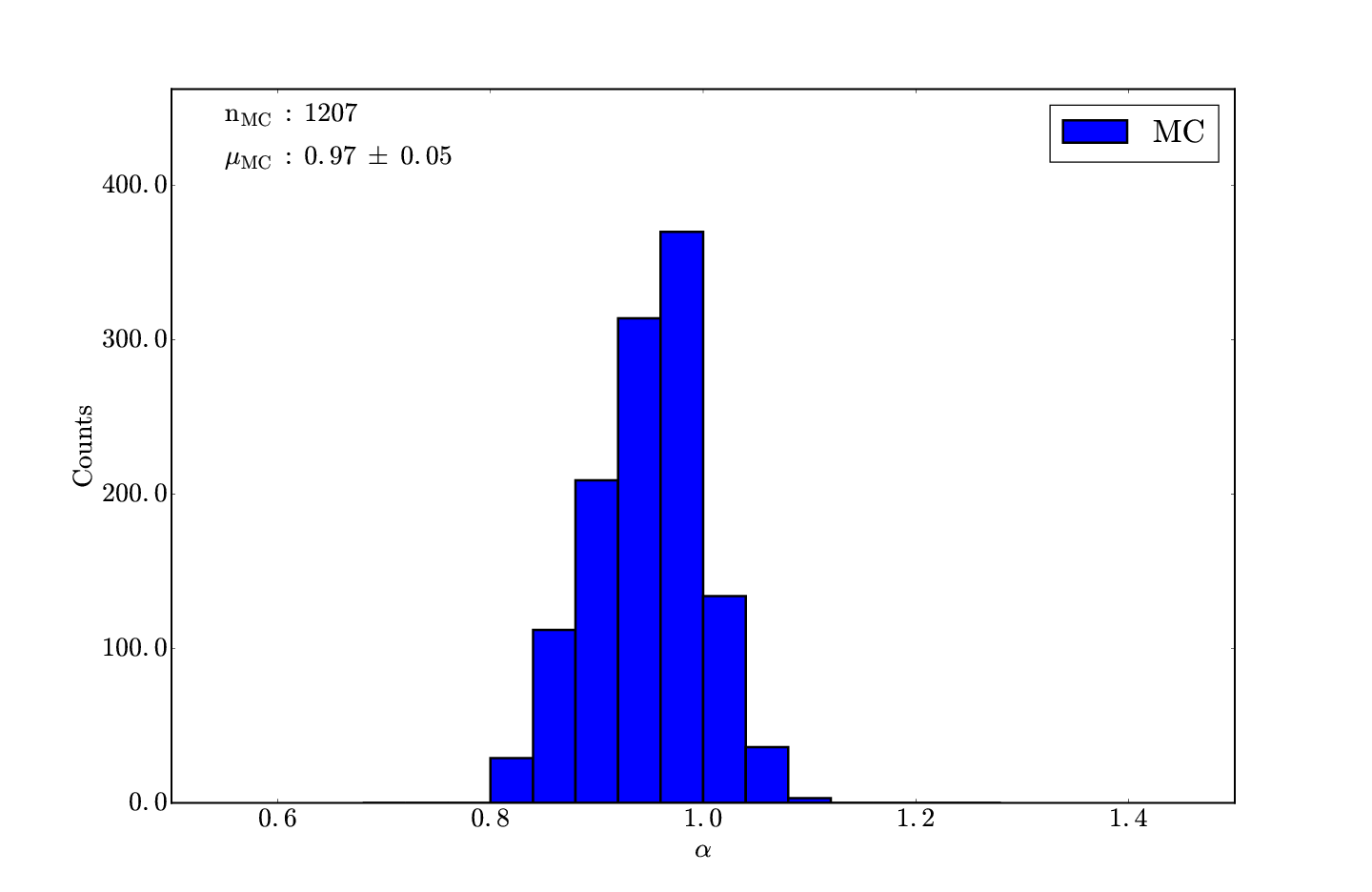}
\else
I am not enabling plots.
\fi
\caption{Histogram of $\alpha$ located at the peak value of $N_{out}(a_c,C,dC,\pv,\alpha)^2$ to $N_{in}(a_c,C,dC,\pv,\alpha)$ in each of the $\sim$1,200 trials repeating the V-shape technique for the Karin family. The mean of the distribution is centered at $\alpha$ = 0.97 $\pm$ 0.05 and the bin size in the histogram is 0.04 consistent with \citet{Nesvorny2002b}.}
\label{fig.KarinMC}
\end{figure}

\begin{figure}
\centering
\hspace*{-1.1cm}
\ifincludeplots
\includegraphics[scale=0.3725]{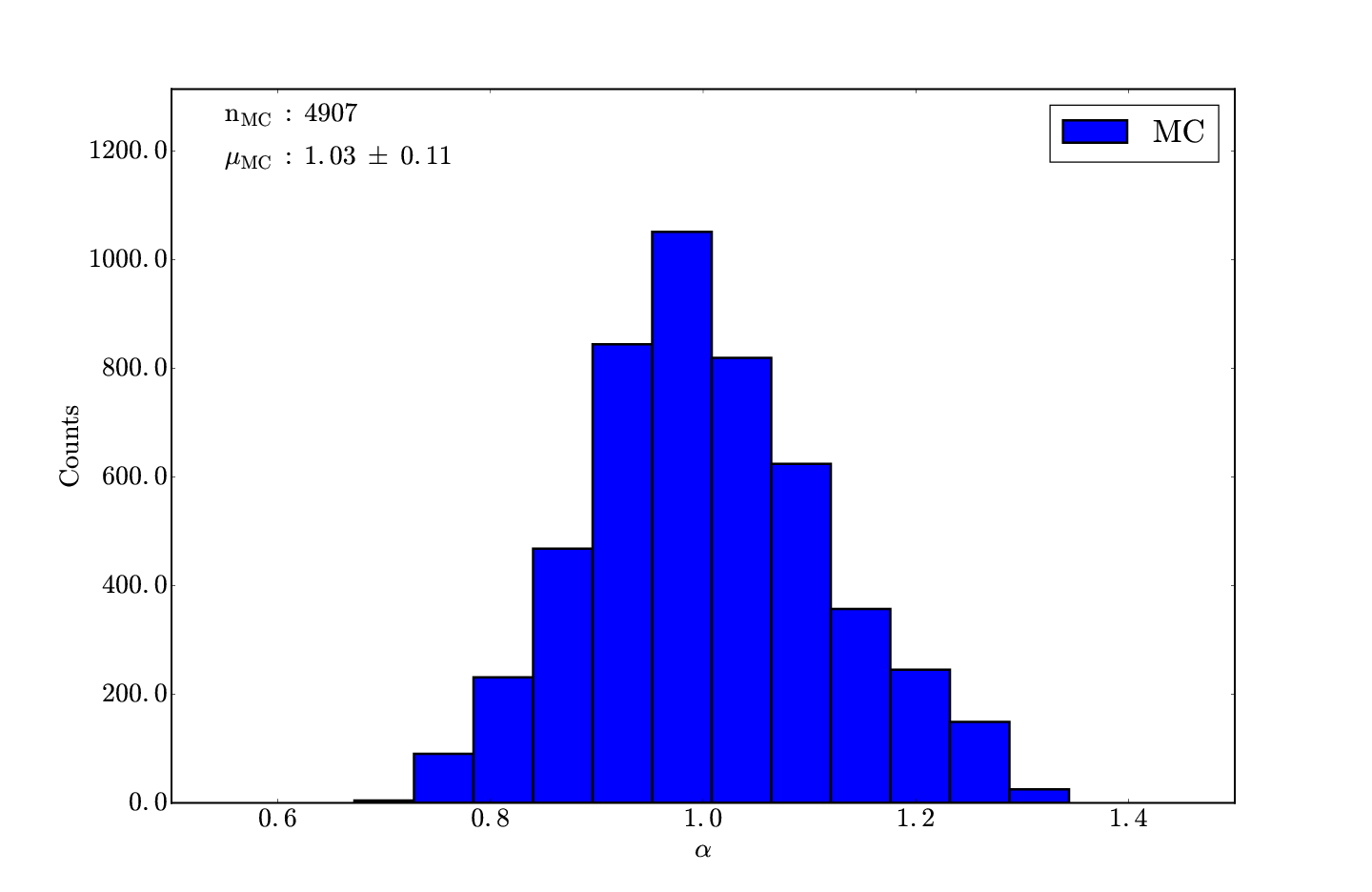}
\else
I am not enabling plots.
\fi
\caption{The same as Fig.~\ref{fig.KarinMC} with $\sim$4,900 trials repeating the V-shape technique for the \FY family. The mean of the distribution is centered at $\alpha$ = 1.03 $\pm$ 0.11 and the bin size in the histogram is 0.06.}
\label{fig.1993fy12MC}
\end{figure}

The mean and root mean square (RMS) uncertainty of $\alpha$  was determined from the distribution of $\alpha$ in the Monte Carlo trials. 
Having more fragments and a well defined-V-shape causes the Monte Carlo technique to produce a narrower distribution in $\alpha$ (E.g.,~ for the Karin family, $\alpha \; = \; 0.97 \pm 0.05$, Fig.~\ref{fig.KarinMC}), while having fewer fragments and a more diffuse V-shape  results in a broader $\alpha$ distribution (\eg,~for the \FY \ family, $\alpha \; = \; 1.03 \pm 0.11$, Fig.~\ref{fig.1993fy12MC}).

\section{Results}
\label{s.results}
\subsection{Synthetic family}
\label{s.synfamily}
It is generally expected that $\alpha_{EV} \; \simeq$ 1 as discussed in Section~\ref{s.Vshapedefinition}. Recent work on the V-shapes of asteroid families $>$100 Myrs-old suggests that $\alpha_{YE} \; \simeq$ 0.8 \citep[][ ]{Bolin2017b}. The time it takes to transition from a V-shape having its $\alpha$ equal to  $\alpha_{EV} \; \simeq \; 1.0$ to $\alpha$ equal to $\alpha_{YE} \; \simeq \; 0.8$ is the time it takes for families to have their initial ejection velocity fields erased by the Yarkovsky effect. We determine the time when the Yarkovsky effect erases $\alpha_{EV}$ by simulating the initial ejection of fragments from the disruption of a parent body and their subsequent spreading caused by the Yarkovsky effect.

The break up of a synthetic asteroid family and its fragments' subsequent evolution due to the Yarkovsky effect is simulated by using 650 particles at $\left (a ,e , \sin i \right ) \; = \; \left (2.37, 0.21, 0.08 \right )$ and distributed in $a$ vs. $D_r$ space according to Eq.~\ref{eqn.deltaaejectionD_r_theta_vs_a_c} with $\alpha_{EV} \; = \; 1.0$ and $V_{EV}$ = 30 $\mps$ using fragments with $2 \mathrm{km} \; < \; D \; < \; 75 \mathrm{km}$ distributed according to the known members of the Erigone family defined by \citet{Nesvorny2015a}.  The eccentricity and inclination distributions were determined by using Gaussian scaling described in \citet[][]{Zappala2002}. $V_{EV}$ = 30 $\mps$ corresponds to a typical initial displacement of $\sim 7.0 \times 10^{-3}$ au for a 5 km diameter asteroid. The Yarkovsky drift rates were defined with
\begin{equation}
\label{eqn.yarkorate_final}
\frac{da}{dt} (D,\alpha,a,e,N,A) \; =\;  \left ( \frac{da}{dt} \right )_{0} \; \frac{\sqrt{a}_{0}(1-e^2_{0})}{\sqrt{a}(1-e^2)} \; \left ( \frac{D_0}{D} \right )^{\alpha_{YE}} \; \left ( \frac{\rho_0}{\rho} \right ) \; \left ( \frac{1 - A}{1 - A_{0}} \right )\; \left ( \frac{\au}{\myr} \right ) \; \frac{cos(\theta)}{cos(\theta_0)}
\end{equation} 
from \citet[][]{Spoto2015,Bolin2017b} with $\left ( \frac{da}{dt} \right )_{0} $ $\sim$ 4.7 $\times \; 10^{-5} \; \au \; \myr^{-1}$, $a_0 \; = \; $ 2.37 au, $e_0 \; = \; $ 0.2, $D_0$ = 5 km, $\rho_0$  = 2.5 $\gpcmc$ and bond albedo, $A_0$, is equal to 0.1, surface conductivity between 0.001 and 0.01 $\mathrm{W \; m^{-1} \; K^{-1}}$ and $\theta_0 \; =$ 0$\tdeg$ \citep[][]{Bottke2006,Vokrouhlicky2015}. For the synthetic family, $\rho$ = 2.3 $\gpcmc$, $A \; = \; 0.02$ and $cos (\theta)$ is uniformly distributed between -1 and 1. An $\alpha_{YE}$ = 0.8 was chosen as $\alpha$ measurements of asteroid families old enough to have their fragments significantly modified by the Yarkovsky effect have 0.7 $\lesssim$ $\alpha$ $\lesssim$ 0.9 \citep[][]{Bolin2017b}. The particles were evolved with the Yarkovsky effect and gravitational perturbations from Mercury, Venus, Earth, Mars, Juputer and Saturn using the $SWIFT$\_$RMVS$ code \citep[][]{Levison1994}. Particles are removed from the simulation if they collide with one of the planets or evolve on to small perihelion orbits. YORP rotational and spin-axis variation are not included in the simulation.

 The V-shape identification technique was applied on the synthetic family data at Time = 0 by using the techniques in Section~\ref{s.v-shapeidentificationandalpha}. Eqs.~\ref{eq.border_method_N_outer} and \ref{eq.border_method_N_inner} are integrated using the interval ($-\infty$,$\infty$) for the Dirac delta function $\delta(a_{j}-a)$ and the interval [$0.04,0.60$] for the Dirac delta function $\delta(D_{r,j}-D_r )$. Eq.~\ref{eqn.apvDvsCfinal} is truncated to 0.04 for $D_r$ $ <$ $0.04$ and to 0.60 for $D_r$ $>$ 0.60 . Asteroids with 0.04 $< \; D_r \; <$ 0.60 were chosen because the number of asteroids in this $D_r$ is large enough so that the leading edge of the V-shape is defined by asteroids with $\mathrm{cos}(\theta) \; = \; $1.0 or -1.0 according to Eq.~\ref{eqn.deltaaejectionD_r_theta_vs_a_c}. 
 
 The V-shape identification technique located a peak at $(a_c, \; C, \; \alpha) \; = \; (2.366 \; \mathrm{au}, \; 2.4 \times 10^{-6} \; \mathrm{au}, \;  \sim1.0)$ as seen in the top panel of Fig.~\ref{fig.synErig0Myrs}. The peak value of $\frac{N_{in}^2}{N_{out}}$ is  $\sim$11 standard deviations above the mean value of $\frac{N_{in}^2}{N_{out}}$ in the range 2.35 au $< \; a \;<$ 2.38 au, 1.0 $\times \; 10^{-6}$ au $< \; C \;<$ 6.0 $\times \; 10^{-6}$ au and 0.8 $< \; \alpha \;<$ 1.2.  A $dC \; = \; 4.0 \; \times 10^{-7}$ au was used. The concentration of the peak to one localized area in $\alpha$ vs. $C$ space is due to the sharpness of the synthetic family's V-shape border.
 
\begin{figure}
\centering
\hspace*{-0.7cm}
\ifincludeplots
\includegraphics[scale=0.425]{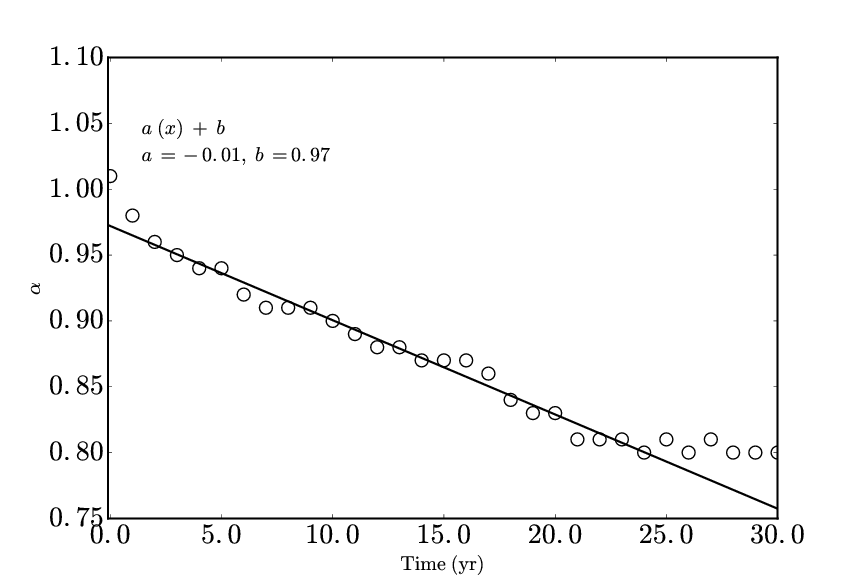}
\else
I am not enabling plots.
\fi
\caption{Time vs. $\alpha$ for the first 30 Myrs of time evolution of the fragments of a synthetic asteroid family given an $\alpha_{EV} \; = \; $1.0  according to Eq.~\ref{eqn.deltaaejectionD_r_theta_vs_a_c} and evolved in time $\alpha_{YE} \; = \; $0.8 according to Eq.~\ref{eqn.yarkorate_final}. The steep decrease in $\alpha$ after $\sim$0 Myrs is due to the randomization of asteroid fragment obliquity and subsequent evolution of their $a$'s according to Eq.~\ref{eqn.yarkorate_final}. The dark line is a linear fit to the first 20 Myrs of simulation data.}
\label{fig.timevsalpha}
\end{figure}

 The V-shape identification technique was applied to family fragments at 1 Myr steps for the first 100 Myrs of the simulation. The value of $\alpha$ for the V-shape linearly decreases below 1.0 and reaches $\sim$0.8, equal to $\alpha_{YE}$ after $\sim$ 20 Myrs (see Fig.~\ref{fig.timevsalpha}). There is a steep drop in $\alpha$ from $\sim$1.0 to $\sim$0.97 in the first 1-2 Myrs of the simulation that is possibly due to the fragments near the borders of the family becoming spread in $a$ according to Eq.~\ref{eqn.yarkorate_final}. This is because fragments at the border of the family with cos($\theta$) = $\pm$1 are the Yarkovsky front-runners and cause a very quick spreading of the family, rapidly changing the value of $\alpha$. Assuming $\alpha_{YE}$ equal to $\sim$0.8 for the Yarkovsky drift size dependence in the Main Belt, asteroid family V-shapes with a measured $\alpha$ closer to 0.8 reveal that the dispersion of the family is dominated by the Yarkovsky effect over the initial ejection of fragments and older than $\sim20$ Myrs. Family V-shapes that have measured $\alpha$ values of $\sim$1.0 have the contribution of the initial ejection velocity field dominant to the value of $\alpha$ and the measured value of $\alpha$ is characteristic of the size-dependece of the initial ejection velocity.

\subsection{Young asteroid family ejection velocity V-shapes}
The V-shape identification technique was applied to 11 asteroid families noted for their young ages between a few Myrs to a few 10 Myrs \citep[][]{Spoto2015,Nesvorny2015a} listed in the first column of Table~\ref{t.youngfams}, selected for having $C-C_{EV} \lesssim0$ as explained in Section~\ref{s.yarkoage}. The measured value of their V-shape's $\alpha$ and uncertainties determined by the techniques in Sections~\ref{s.v-shapeidentificationandalpha} and \ref{s.montecarlo} as well as physical properties used to measure the $\alpha$ of family V-shapes are summarized in Table~\ref{t.youngfams}. A description of how the V-shape identification technique is implemented for each family is described in the Appendix.

\begin{table}[]
\centering
\caption{Description of variables in order of appearance.}
\label{my-label}
\begin{tabular}{ll}
\hline
\multicolumn{1}{|l|}{Variable}      & \multicolumn{1}{l|}{Description}                                                      \\ \hline
\multicolumn{1}{|l|}{$D$}           & \multicolumn{1}{l|}{Asteroid diameter in km}                                                                 \\ \hline
\multicolumn{1}{|l|}{$a$}           & \multicolumn{1}{l|}{Semi-major axis in au.}                                           \\ \hline
\multicolumn{1}{|l|}{$e$}           & \multicolumn{1}{l|}{Eccentricity.}                                                    \\ \hline
\multicolumn{1}{|l|}{$i$}           & \multicolumn{1}{l|}{Inclination in degrees.}                                                     \\ \hline
\multicolumn{1}{|l|}{$D_r$}         & \multicolumn{1}{l|}{Reciprocal of the diameter, $\frac{1}{D}$ in km$^{-1}$.}          \\ \hline
\multicolumn{1}{|l|}{$a_c$}         & \multicolumn{1}{l|}{The location of the V-shape centere in au.}                        \\ \hline
\multicolumn{1}{|l|}{$n$}           & \multicolumn{1}{l|}{Mean motion in $\frac{\mathrm{rad}}{\mathrm{s}}$}             \\ \hline
\multicolumn{1}{|l|}{$V_{ev}$}      & \multicolumn{1}{l|}{Ejection velocity in $\frac{\mathrm{m}}{\mathrm{s}}$.}            \\ \hline
\multicolumn{1}{|l|}{$\alpha_{EV}$} & \multicolumn{1}{l|}{The $\alpha$ of an initial velocity V-shape.}                               \\ \hline
\multicolumn{1}{|l|}{$\pv$}         & \multicolumn{1}{l|}{Visual albedo.}                                                   \\ \hline
\multicolumn{1}{|l|}{$C$}           & \multicolumn{1}{l|}{Total V-shape width in au.}                                       \\ \hline
\multicolumn{1}{|l|}{$\alpha$}      & \multicolumn{1}{l|}{$\alpha$ defined by Eq.~\ref{eqn.apvDvsCfinal}.}           \\ \hline
\multicolumn{1}{|l|}{$N_{out}$}     & \multicolumn{1}{l|}{Number density of objects between the nominal and outer V-shapes.} \\ \hline
\multicolumn{1}{|l|}{$N_{in}$}      & \multicolumn{1}{l|}{Number density of objects between the nominal and inner V-shapes.} \\ \hline
\multicolumn{1}{|l|}{$dC$}          & \multicolumn{1}{l|}{Difference in $C$ between the nominal and outer/inner V-shapes.}  \\ \hline
\multicolumn{1}{|l|}{$H$}           & \multicolumn{1}{l|}{Absolute magnitude.}                                              \\ \hline
\multicolumn{1}{|l|}{$C_{YE}$}      & \multicolumn{1}{l|}{V-shape width due to Yarkovsky spreading of fragments in au.}     \\ \hline
\multicolumn{1}{|l|}{$C_{EV}$}      & \multicolumn{1}{l|}{V-shape width due to the initial ejection of fragments in au.}    \\ \hline
\multicolumn{1}{|l|}{$N$}           & \multicolumn{1}{l|}{Number of family members used with the V-shape technique.}        \\ \hline
\multicolumn{1}{|l|}{$\alpha_{YE}$} & \multicolumn{1}{l|}{The $\alpha$ of a Yarkovsky V-shape.}                                  \\ \hline
\multicolumn{1}{|l|}{$\rho$}        & \multicolumn{1}{l|}{Asteroid density in $\gpcmc$.}                                    \\ \hline
\multicolumn{1}{|l|}{$A$}           & \multicolumn{1}{l|}{Bond albedo.}                                                     \\ \hline
\multicolumn{1}{|l|}{$\theta$}      & \multicolumn{1}{l|}{Asteroid obliquity.}                                              \\ \hline
                                    &                                                                                      
\end{tabular}
\end{table}

\begin{table}[]
\centering
\hspace*{-1.35cm}
\label{t.youngfams}
\begin{tabular}{|l|l|l|l|l|l|l|l|l|l|}
\hline
Designation & Tax. & D$_{pb}$& $C_{EV}$& $C$ &$N$ & $a_{c}$ & $\alpha$ & $\pv$ & $D_{s}$ - $D_{l}$  \\ \hline
 &  & (km) & ($10^{-6}$ au) & ($10^{-6}$ au) &  & (au)   &  &   & (km) \\ \hline
\FY       &    S  &      15         &7.5  &8.1    &    87               &        2.847            & 1.03 $\pm$ 0.11&    0.17 $\pm$ 0.05   &     2.0 - 4.7                \\ \hline
Aeolia       &    X  &    35   &7.7 &3.6  &  225                       &       2.7415              & 1.0 $\pm$ 0.07 &    0.11 $\pm$ 0.03   &    1.3 - 3.3                \\ \hline
Brangane    & S     &        42       & 7.2 & 3.9 &            171           &    2.584                 & 0.95 $\pm$ 0.04 &    0.1 $\pm$ 0.03  &      1.5 - 7.5               \\ \hline
Brasilia    &   X   &      34         & 13.4 & 15.2 &                   548  &     2.855                &  1.0 $\pm$ 0.1        &    0.24 $\pm$ 0.07   & 1.0 - 7.0                    \\ \hline
Clarissa    &   C   &      39         & 3.9 & 3.8 &                 179  &     2.404                &  0.95 $\pm$ 0.04        &    0.06 $\pm$ 0.02   & 1.1 - 8.9                    \\ \hline
Iannini       &    S  &      14$^{*}$         &  5.6 &  2.2 &    129                  &               2.644     & 0.97 $\pm$ 0.07 &    0.30 $\pm$ 0.10   & 0.9 - 3.7                    \\ \hline
Karin       &    S  &      40         & 12.1 &3.1 &    429               &         2.865            & 0.97 $\pm$ 0.05 &    0.21 $\pm$ 0.06   &           1.8 - 10.3          \\ \hline
K{\"o}nig       &    C  &      37         & 3.3 &  4.4   & 315                   &             2.574        &  0.91 $\pm$ 0.03  &    0.06 $\pm$ 0.01   &          1.4 - 5.5           \\ \hline
Koronis(2)       &    S  &      58         & 11.1 &   2.3   &    235               &         2.869            & 1.09 $\pm$ 0.05 &    0.14 $\pm$ 0.04   &           1.1 - 3.6          \\ \hline
Theobalda       &    C  &      97        & 10.7 & 9.0   &  349                 &      3.179           & 0.95 $\pm$ 0.04  &    0.06 $\pm$ 0.02   &    2.1 - 15.2             \\ \hline
Veritas       &    C  &      124$^{*}$         & 21.0 & 12.2   & 1135                  &               3.168     & 1.01 $\pm$ 0.04 &    0.07 $\pm$ 0.02   & 3.2 - 31.7                    \\ \hline

\end{tabular}
\caption{The measured value of $\alpha$ of young asteroid families: Asteroid family fragment taxonomies are taken from \citet[][]{Nesvorny2003, Willman2008, Harris2009, Molnar2009, Novakovic2010, Spoto2015, Nesvorny2015a}. Diameters for the parent body, $D_{pb}$, were taken from the means of asteroid family parent bodies in \citet[][]{Broz2013b} and \citet[][]{Durda2007} if $D_{pb}$ was available from both sources. $D_{pb}$ for Iannini was taken from \citet[][]{Nesvorny2003}. The $D_{pb}$ of the Koronis(2)and Veritas families were determined with techniques from \citet[][]{Nesvorny2015a} to estimate parent body size. $\pv$ values of asteroid family members are taken from \citet[][]{Masiero2013} and \citet[][]{Spoto2015}. $N$ is the number of family fragments used in the determination of a family's V-shape. $D_s$ and $D_l$ are the boundaries of the $D$ of the smallest and largest fragments used to measure an asteroid family v-shape's $\alpha$.}
\end{table}

\section{Discussion and Conclusion}

We have demonstrated that the techniques of \citet[][]{Bolin2017} can be used not only to identify asteroid family V-shapes, but also to measure the spreading of family fragments caused by the initial ejection velocity field from the disruption of the parent body and by the Yarkovsky effect. We have demonstrated, following the work of \citet[][]{Vokrouhlicky2006b}, that the functional form of the spread of a family created entirely initial ejection field of fragments from their parent body's disruption (\ie with $C_{YE}$ = 0 in Eq.~\ref{eqn.Ccombo}), and the spread caused by the Yarkovsky effect (\ie with $C_{EV}\; =$ 0 and $C_{YE}\; >$  0 in Eq.~\ref{eqn.Ccombo}) are functionally equivalent. 

We have measured the V-shapes of 11 young ($<$100 My-old) asteroid families located within the inner, central and outer MB and we have found that all of them $\alpha \simeq$1.0. We associate this value of $\alpha$ to the initial velocity V-shape's $\alpha_{EV}$, concluding that the initial ejection velocity is proportional to $\frac{1}{D}$, as was already assumed, because these families are too young to have been substantially modified by the Yarkovsky effect. Our $\alpha$ measurements were repeated for each of the 11 families using the Monte Carlo scheme described in Section~\ref{s.montecarlo} and found that the 1 $\sigma$ uncertainty of the trial $\alpha$ measurements were within $\sim5\%$ of the mean trial $\alpha$ value for most of the families. Some of the families such as the Aeolia, Clarissa, Brangane, Iannini and Koronis(2) families had slightly skewed trial $\alpha$ distributions in the positive and negative directions.

The average value of $\alpha$ of the 11 family V-shapes in this study is 0.97 $\pm$ 0.02, or within in the Student's t-distribution 99.8$\%$ confidence interval  0.95 - 0.99. Additionally, our average measurement of $\alpha \; \simeq$ 1.0 confirms the results of laboratory and numerical experiments of asteroid disruption events showing $\alpha_{EV} \; \simeq$ 1.0 \cite[][]{Fujiwara1989,Michel2001, Nesvorny2004}. Other studies focusing on modelling the observed distribution of family fragments with disruption simulations such as for the Karin family \citep[][]{Nesvorny2002b, Nesvorny2006a} or the $i$ distribution of family such as for the Koronis family \citep[][]{Carruba2016e} also show that $\alpha \; \simeq$ 1.0. 

The $\alpha$ determination technique in Section~\ref{s.v-shapeidentificationandalpha} can be used to determine whether or not an asteroid family V-shape is young enough to not have been significantly altered by the Yarkovsky effect. There is indication that the value of $\alpha$ for Yarkovsky V-shapes, or $\alpha_{YE}$, for older families whose fragments have been significantly modified by the Yarkovsky effect as described by Eq.~\ref{eqn.apvDvsCfinal} is between 0.7 - 0.9 due to possible thermal inertia dependence on asteroid size and its effect on the Yarkovsky drift rate as a function of asteroid size \citep[][]{Delbo2007, Delbo2015, Bolin2017b}. Asteroid family V-shapes that have been significantly affected by the Yarkovsky effect will have $\alpha$ values as described in Eq.~\ref{eqn.apvDvsCfinal} that are closer to inside the range of 0.7 - 0.9.

In fact, the ages of some of these families have been determined by the use of alternative methods such as backwards integrating the orbits of selected bodies in the families or by modedlling the diffusion of fragments caused by chaos \cite[][]{Nesvorny2002b,Nesvorny2003,Tsiganis2007,Novakovic2010,Carruba2017a}. By these methods, Iannini, Karin, Theobalda and Veritas all have ages between 5 and 9 Myrs independently ruling out significant modification of the Yarkovsky effect on their fragments' $a$.

We demonstrate that V-shapes in $a$ vs $D_r$ space created purely by the initial ejection of fragments can be separated from those that are created by a combination of the initial ejection velocity of fragments and the Yarkovsky effect using the measurement of $\alpha$ as generically described for asteroid family V-shapes by Eq.~\ref{eqn.apvDvsCfinal}.  We assume that the value of $\alpha_{YE}$ $\sim$0.8 as indicated as indicated by \citet[][]{Bolin2017b} for very old families which have lost memory of their initial dispersion. We find that the time-scale for $\alpha$ to reach $\alpha_{YE}$ as a result of the modification of fragments' $a$ by the Yarkovsky effect to is on the order of $\sim$20 Myrs as seen in Fig.~\ref{fig.timevsalpha}. Interestingly, the backwards integration technique is unable to identify families and determine accurate family ages for families older than $\sim$20 Myrs \citep[][]{Nesvorny2003, Radovic2017b}. 

The V-shape identification method measuring family V-shapes' $\alpha$ provides a way to distinguish whether a family V-shape is caused by a combination of the initial ejection of fragments and the Yarkovsky effect, or only due to initial ejection of fragments. The measurement of asteroid family V-shape provides an additional independent evidence of the subsequent orbital evolution of asteroid family fragment due to the Yarkovsky effect after the initial placement of fragments due to the initial ejection of fragments because asteroid families old enough to have their fragments' orbits modified by the Yarkovsky will have an $\alpha \; <$ 1.0 compared to the case where asteroid family fragments' orbits remain unmodified after their parent's disruption where their family V-shape would have $\alpha$ = 1.0.

It may be possible to independently constrain the degree to which an asteroid family fragments' spread in $a$ has been modified in part by the Yarkovsky effect relative to the spread caused by the initial ejection of fragments. Family V-shapes with a higher proportion of $C_{EV}$ relative to $C_{YE}$ in their total value of $C$ may have a higher $\alpha \; <$ 1 compared to family V-shapes with a higher proportion of $C_{YE}$ relative to $C_{EV}$ in their total value of $C$. Distinguishing the families with higher proportion of $C_{EV}$ compared to $C_{YE}$ requires measurements of $\alpha$ with small uncertainties. Using methods of removing outliers by colours and other physical data such as the method of \citet[][]{Radovic2017b} may improve the precision of $\alpha$ measurements independently of other methods such as the V-shape criterion of \citet[][]{Nesvorny2015a} which may bias the measurement of a V-shape's $\alpha$ towards the assumption of $\alpha$ used in the V-shape criterion before the actual measurement of $\alpha$ is made. Understanding more about the evolution of family fragments' $a$ and its affect on the measurement of $\alpha$ may provide an independent constraint on a family's age provided the physical properties of the asteroid family fragments are known such as density and surface thermal conductivity. 

% ACKNOWLEDGEMENTS -----------------------------------------------------------

\acknowledgments

\section*{Acknowledgments}

We would like to thank the reviewer of our manuscript, Valerio Carruba, for providing helpful comments and suggestions for improving the quality of the text. BTB is supported by l'\`{E}cole Doctorale Sciences Fondatementales et Appliqu\'{e}es, ED.SFA (ED 364) at l'Universit\'{e} de Nice-Sophia Antipolis. KJW was supported by the National Science Foundation, Grant 1518127. BTB would like to acknowledge James W. Westover for thought-provoking discussions on the implementation of large-scale computing resources and algorithms that were used in the completion of this work.

\appendix
\section{Appendix}
\label{s.Appendix}

\subsection{\FY}
\label{s.FY}

\begin{figure}
\centering
\hspace*{-0.7cm}
\ifincludeplots
\includegraphics[scale=0.425]{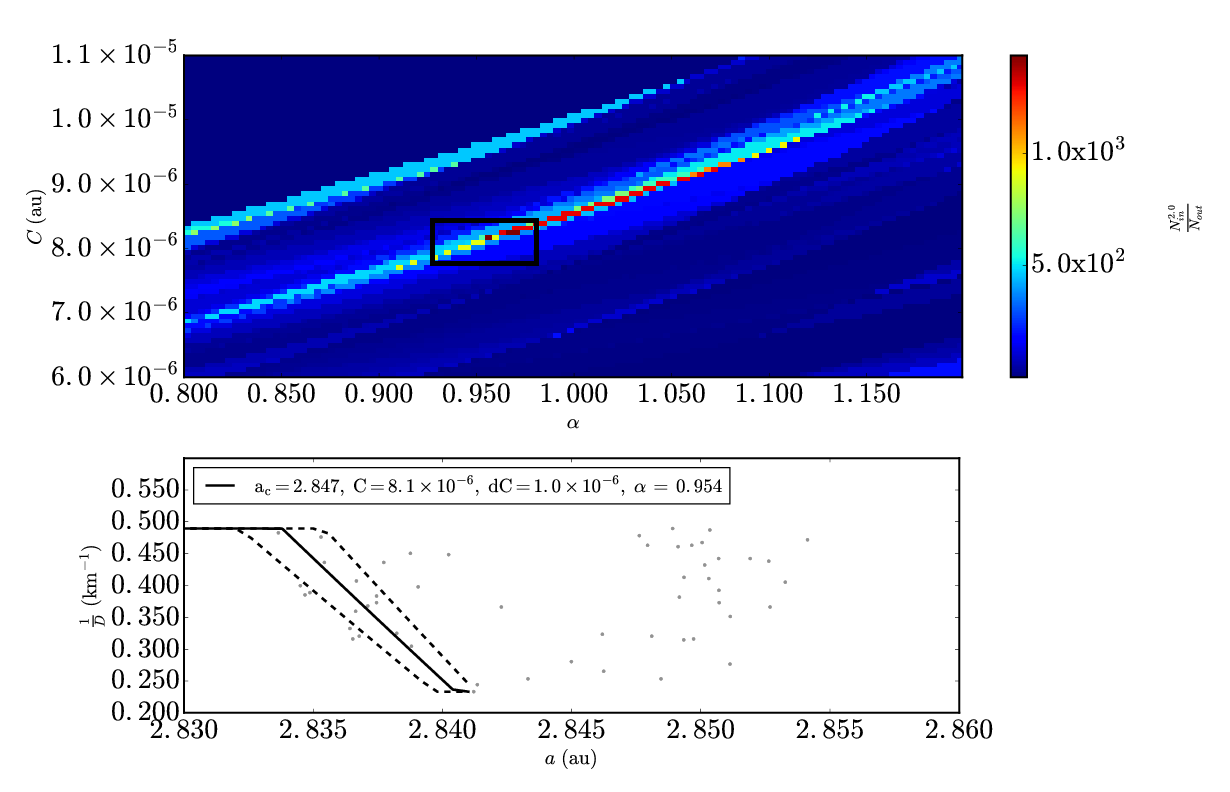}
\else
I am not enabling plots.
\fi
\caption{The same as Fig.~\ref{fig.synErig0Myrs} for \FY asteroid family data from \citet[][]{Nesvorny2015a}. (Top panel) $\Delta \alpha$ is equal to $7.0 \times 10^{-3}$ au and $\Delta C$, is equal to $1.5 \times 10^{-7}$ au. (Bottom Panel) $D_r(a,a_c,C\pm dC,\pv,\alpha)$ is plotted with $\pv = 0.17$. $a_c$ = 2.847 au and $dC \; = \; 8.0 \x 10^{-7}$ au.}
\label{fig.1993fy12BorderAlph}
\end{figure}

The \FY asteroid family located in the outer Main Belt was first identified by \citet[][]{Nesvorny2010a} and consists of mostly S-type asteroids \citep[][]{Spoto2015}. The age of the family is roughly estimated to be $<$200 Myrs by \citet[][]{Broz2013b} and $83 \pm 28$ Myrs by \citet[][]{Spoto2015}. It should be noted that ages of asteroid families by \citet[][]{Spoto2015} are upper limits to the family age because they are computed with the assumption that $C_{EV} \; \simeq$0. The V-shape identification technique was applied to 87 asteroids belonging to the \FY asteroid family defined by \citet[][]{Nesvorny2015a}. Eqs.~\ref{eq.border_method_N_outer} and \ref{eq.border_method_N_inner} are integrated using the interval ($-\infty$,$a_c$] for the Dirac delta function $\delta(a_{j}-a)$ because the inner half of the family V-shape for \FY  is more densely populated than the outer V-shape half possibly due to lack of completeness of \FY family members as seen in the bottom panel of Fig.~\ref{fig.1993fy12BorderAlph}. The interval [$0.21,0.49$] for the Dirac delta function $\delta(D_{r,j}-D_r )$ to avoid the potential distortion of the family V-shape because of smaller fragments interacting with the 5:2 MMR at 2.81 au.  Eq.~\ref{eqn.apvDvsCfinal} is truncated to 0.21 for $D_r$ $ <$ $0.21$ and to 0.49 for $D_r$ $>$ 0.49. Asteroid $H$ values were converted to $D$ using Eq.~\ref{eq.HtoD} using the value of $\pv$ = 0.184 typical for members of the \FY family \citep[][]{Spoto2015}.

\clearpage
    
 The peak in $\frac{N_{in}^2}{N_{out}}$ at $(a_c, \; C, \; \alpha) \; = \; (2.847 \; \mathrm{au}, \; 8.1 \times 10^{-6} \; \mathrm{au}, \;  \sim0.95)$ as seen in the top panel of Fig.~\ref{fig.1993fy12BorderAlph} is located in the range 2.83 au $< \; a \;<$ 2.86 au, 6.0 $\times \; 10^{-6}$ au $< \; C \;<$ 1.1 $\times \; 10^{-5}$ au and 0.8 $< \; \alpha \;<$ 1.2.  A $dC \; = \; 1.0 \; \times 10^{-6}$ au was used. The technique was repeated with the \FY family defined by \citet[][]{Spoto2015} resulting in identical results.
 
 The V-shape identification technique was repeated in $\sim$4,900 Monte Carlo runs where $H$ magnitudes were randomised by the typical magnitude uncertainty of 0.25 for asteroids in the MPC catalogue \citep[][]{Oszkiewicz2011,Veres2015} and their $\pv$ was assumed to be the average value of $\pv$ for family fragments in the \FY family fragments of 0.18 with an uncertainty of 0.04 \citep[][]{Spoto2015}. The value of $\alpha$ in $\sim$4,900 Monte Carlo trials ranges between 0.5 $\lesssim$ $\alpha$ $\lesssim$ 1.5 and is on average $\sim$1.0 with a RMS uncertainty of 0.11 as seen in Fig.~\ref{fig.1993fy12MC}. The large RMS uncertainty of 0.11 is due to the low number of asteroids used to measure the family V-shape's $\alpha$. The value of $\mu_{\alpha} \; \simeq \; $ 1.0 and a similar value of $C\; = \; 8.1 \; \times 10^{-6}$ compared to $C_{EV} \; = \; 7.50 \; \times 10^{-6}$ calculated using Eq.~\ref{eqn.VEVvsCalphaFinal} assuming $V_{EV}$ = 15 $\mps$ from \citet[][]{Broz2013b} suggests that the spread of fragments in the \FY family in $a$ vs. $D_r$ space is almost entirely due to the ejection velocity of the fragments with minimal modification in $a$ due to the Yarkovsky effect.
 
\subsection{Aeolia}
\label{s.Aeolia}

\begin{figure}
\centering
\hspace*{-0.7cm}
\ifincludeplots
\includegraphics[scale=0.425]{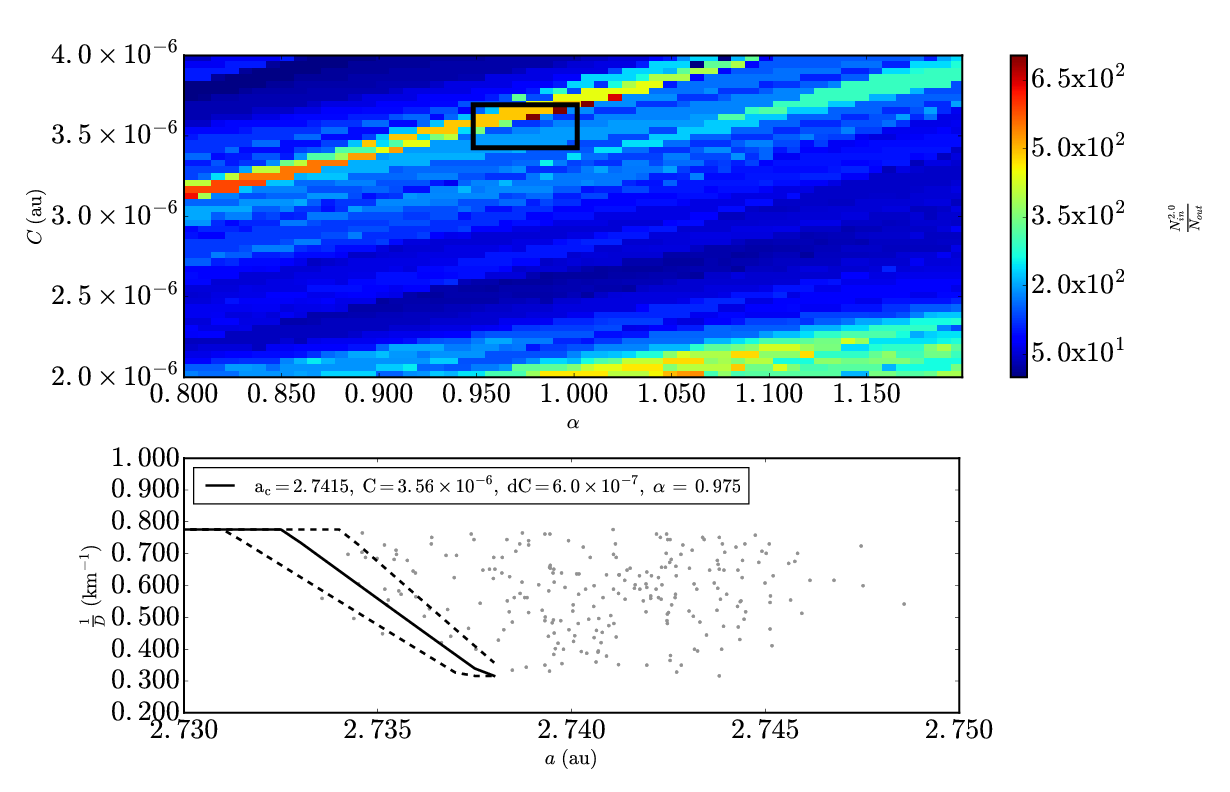}
\else
I am not enabling plots.
\fi
\caption{The same as Fig.~\ref{fig.synErig0Myrs} for Aeolia asteroid family data from \citet[][]{Nesvorny2015a}. (Top panel) $\Delta \alpha$ is equal to $7.0 \times 10^{-3}$ au and $\Delta C$, is equal to $8.0 \times 10^{-8}$ au. (Bottom Panel) $D_r(a,a_c,C\pm dC,\pv,\alpha)$ is plotted with $\pv = 0.11$, $a_c$ = 2.742 au and $dC \; = \; 3.56 \x 10^{-6}$ au.}
\label{fig.AeoliaBorderAlph}
\end{figure}

The X-type Aeolia asteroid family located in the outer Main Belt was first identified by \citet[][]{Nesvorny2010a}. The age of the family is estimated to be only $<$100 Myrs by \citet[][]{Broz2013b} and $100 \pm 18$ Myrs by \citet[][]{Spoto2015}.  The V-shape identification technique was applied to 225 asteroids belonging to the Aeolia asteroid family defined by \citet[][]{Nesvorny2015a}. Eqs.~\ref{eq.border_method_N_outer} and \ref{eq.border_method_N_inner} are integrated using the interval ($-\infty$,$a_c$] for the Dirac delta function $\delta(a_{j}-a)$ because the outer half of the family V-shape for Aeolia has a less defined border than the inner V-shape half due to possible interlopers as seen in the bottom panel of Fig.~\ref{fig.AeoliaBorderAlph}. The V-shape criterion of \citet[][]{Nesvorny2015a} was not used to remove potential interlopers because it assumes a functional form of $\alpha \; = \; 1.0$ in Eq.~\ref{eqn.apvDvsCfinal} which would result in artificially trimming and biasing the family V-shape towards having an $\alpha \; = \; 1.0$. The interval [$0.31,0.78$] for the Dirac delta function $\delta(D_{r,j}-D_r )$ was used to remove at higher $D_r$ values on the inner edge of the family V-shape in the application of the V-shape identification technique.  Eq.~\ref{eqn.apvDvsCfinal} is truncated to 0.31 for $D_r$ $ <$ $0.31$ and to 0.78 for $D_r$ $>$ 0.78. Asteroid $H$ values were converted to $D$ using Eq.~\ref{eq.HtoD} using the value of $\pv$ = 0.11 typical for members of the Aeolia family \citep[][]{Spoto2015}.

 The peak in $\frac{N_{in}^2}{N_{out}}$ at $(a_c, \; C, \; \alpha) \; = \; (2.742 \; \mathrm{au}, \; 3.56 \times 10^{-6} \; \mathrm{au}, \;  \sim0.975)$ as seen in the top panel of Fig.~\ref{fig.AeoliaBorderAlph} is located in the range 2.73 au $< \; a \;<$ 2.75 au, 2.0 $\times \; 10^{-6}$ au $< \; C \;<$ 4.0 $\times \; 10^{-6}$ au and 0.8 $< \; \alpha \;<$ 1.2.  A $dC \; = \; 6.0 \; \times 10^{-7}$ au was used. 
The technique was repeated with the Aeolia family defined by \citet[][]{Spoto2015} resulting in similar results. In addition, the Aeolia family is noted by \citet[][]{Spoto2015} for having an asymetrical V-shape where the outer half the V-shape has a steeper slope. The $a_c$, $C$, $\alpha$ technique was repeated using [$a_c$,$\infty$) for the Dirac delta function $\delta(a_{j}-a)$ resulting in a peak located at $\frac{N_{in}^2}{N_{out}}$ at $(a_c, \; C, \; \alpha) \; = \; (2.742 \; \mathrm{au}, \;2.46 \times 10^{-6} \; \mathrm{au}, \;  \sim0.968)$. A lower value of $C$ of $2.46 \times 10^{-6}$ for the outer V-shape half compared to the value of $C$ of is in $3.56 \times 10^{-6}$ is in agreement with \citet[][]{Spoto2015} for the outer V-shape half of the Aeolia family having a smaller value of $C$.
 
\begin{figure}
\centering
\hspace*{-1.1cm}
\ifincludeplots
\includegraphics[scale=0.3725]{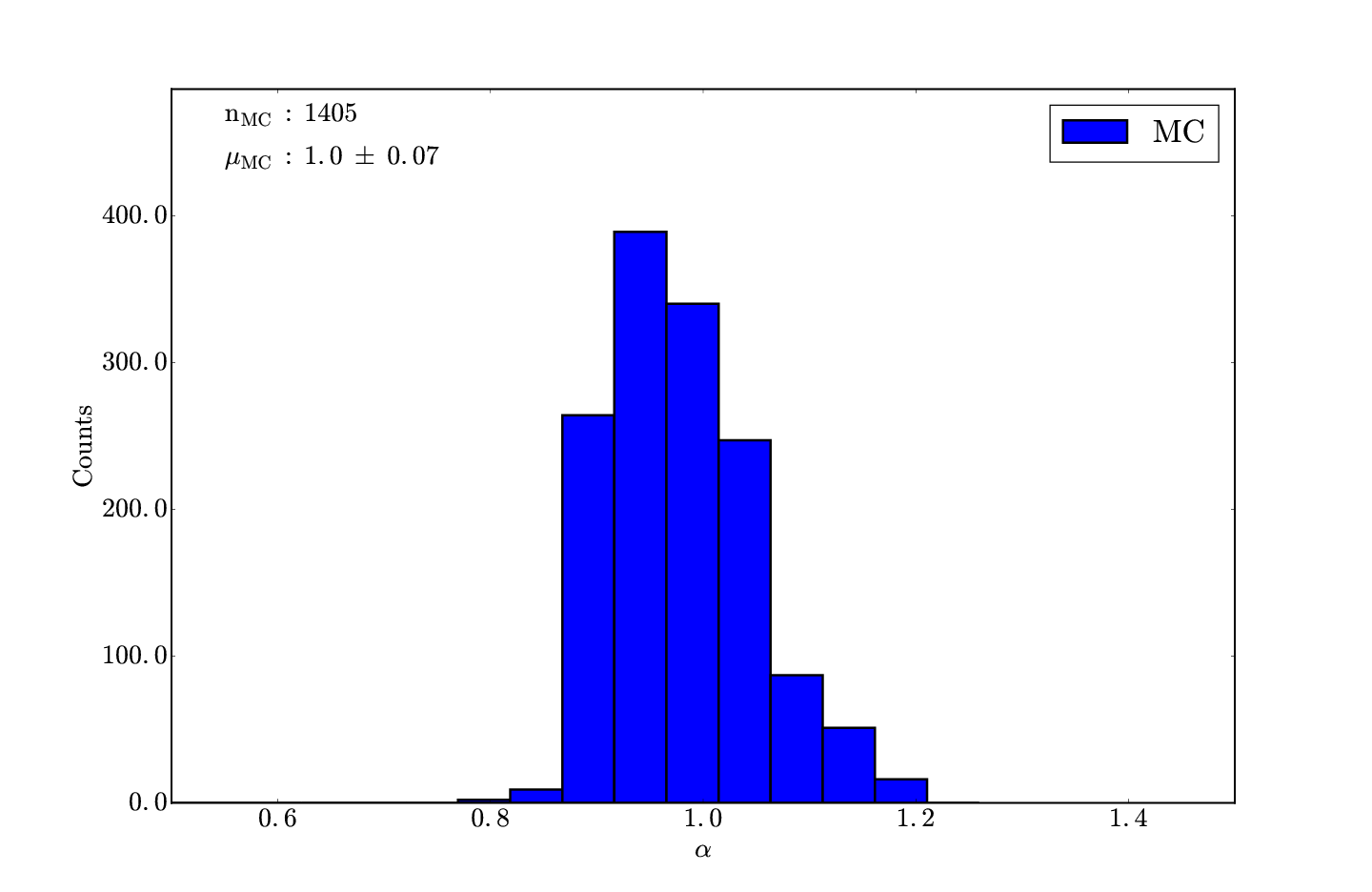}
\else
I am not enabling plots.
\fi
\caption{The same as Fig.~\ref{fig.KarinMC} with $\sim$1,930 trials repeating the V-shape technique for the Aeolia family. The mean of the distribution is centered at $\alpha$ = 1.0 $\pm$ 0.07 and the bin size in the histogram is 0.05.}
\label{fig.AeoliaMC}
\end{figure}

1,405 Monte Carlo runs were completed where $H$ magnitudes were randomised by the typical magnitude uncertainty of 0.25 for asteroids in the MPC catalogue \citep[][]{Oszkiewicz2011,Veres2015} and their $\pv$ was assumed to be 0.11on average with an uncertainty of 0.03 \citep[][]{Spoto2015}. The value of $\alpha$ in 1,405 Monte Carlo trials is on average $\sim$1.0 with a RMS uncertainty of 0.07 as seen in Fig.~\ref{fig.AeoliaMC}. The large RMS uncertainty of 0.07 is due to the low number of asteroids used to measure the family V-shape's $\alpha$ on the inner half of the family V-shape. The $\alpha$ distribution of Monte Carlo runs is slightly positively skewed such that the most probable value is slightly lower than the mean of $\sim1.0$. The value of $\mu_{\alpha} \; \simeq \; $ 1.0 and a smaller value of $C\; = \; 3.6 \; \times 10^{-6}$ au compared to $C_{EV} \; = \; 7.7 \; \times 10^{-6}$ au calculated from Eq.~\ref{eqn.VEVvsCalphaFinal} assuming $V_{EV}$ = 22 $\mps$, the escape velocity from a parent body with a parent body diameter, $D_{pb} \; = \; 35$ km \citep[][]{Durda2007, Broz2013b} and $\rho \; = \; 2.7 \; \gpcmc$ typical for X-type asteroids \citep[][]{Carry2012} suggests that the spread of fragments in the Aeolia family in $a$ vs. $D_r$ space is due to the ejection velocity of the fragments with minimal modification in $a$ due to the Yarkovsky effect.
 
\subsection{Brangane}
\label{s.Brangane}

The S-type Brangane asteroid family located in the central Main Belt was first identified by \citet[][]{Nesvorny2010a}. The age of the family is estimated to be 50$\pm$40 Myrs by \citet[][]{Broz2013b} with a similar estimate in \citet[][]{Spoto2015}. The V-shape identification technique was applied to 171 asteroids belonging to the Brangane asteroid family defined by \citet[][]{Nesvorny2015a}. Eqs.~\ref{eq.border_method_N_outer} and \ref{eq.border_method_N_inner} are integrated using the interval ($-\infty$,$a_c$] for the Dirac delta function $\delta(a_{j}-a)$ because the outer half of the family V-shape for Brangane has a fewer asteroids in $a$ vs, $D_r$ space on its outer V-shape half as seen panel of Fig.~\ref{fig.BranganeAlph}. The interval [$0.19,0.92$] for the Dirac delta function $\delta(D_{r,j}-D_r )$ was used to cover the range of the entire inner half of the Brangane family's V-shape.  Eq.~\ref{eqn.apvDvsCfinal} is truncated to 0.19 for $D_r$ $ <$ $0.19$ and to 0.92 for $D_r$ $>$ 0.92. Asteroid $H$ values were converted to $D$ using Eq.~\ref{eq.HtoD} using the value of $\pv$ = 0.10 typical for members of the Brangane family \citep[][]{Masiero2013, Spoto2015}.
    
The peak in $\frac{N_{in}^2}{N_{out}}$ at $(a_c, \; C, \; \alpha) \; = \; (2.584 \; \mathrm{au}, \; 3.89 \times 10^{-6} \; \mathrm{au}, \;  \sim0.95)$ as seen in the top panel of Fig.~\ref{fig.BranganeAlph} is located in the range 2.57 au $< \; a \;<$ 2.60 au, 3.5 $\times \; 10^{-6}$ au $< \; C \;<$ 5.5 $\times \; 10^{-6}$ au and 0.7 $< \; \alpha \;<$ 1.3.  A $dC \; = \; 1.0 \; \times 10^{-6}$ au was used. The technique was repeated with the Brangane family defined by \citet[][]{Spoto2015} resulting in similar results as seen in Fig.~\ref{fig.BranganeAlphMilani}. The $a_c$, $C$, $\alpha$ technique was repeated using [$a_c$,$\infty$) for the Dirac delta function $\delta(a_{j}-a)$ resulting in a peak located at $\frac{N_{in}^2}{N_{out}}$ at $(a_c, \; C, \; \alpha) \; = \; (2.5835 \; \mathrm{au}, \;3.05 \times 10^{-6} \; \mathrm{au}, \;  \sim0.985)$. A lower value of $C$ of $3.05 \times 10^{-6}$ for the outer V-shape half compared to the value of $C$ of is in $3.83 \times 10^{-6}$ is in agreement with \citet[][]{Spoto2015} for the outer V-shape half of the Brangane family having a smaller value of $C$.
  
\begin{figure}
\centering
\hspace*{-1.1cm}
\ifincludeplots
\includegraphics[scale=0.3725]{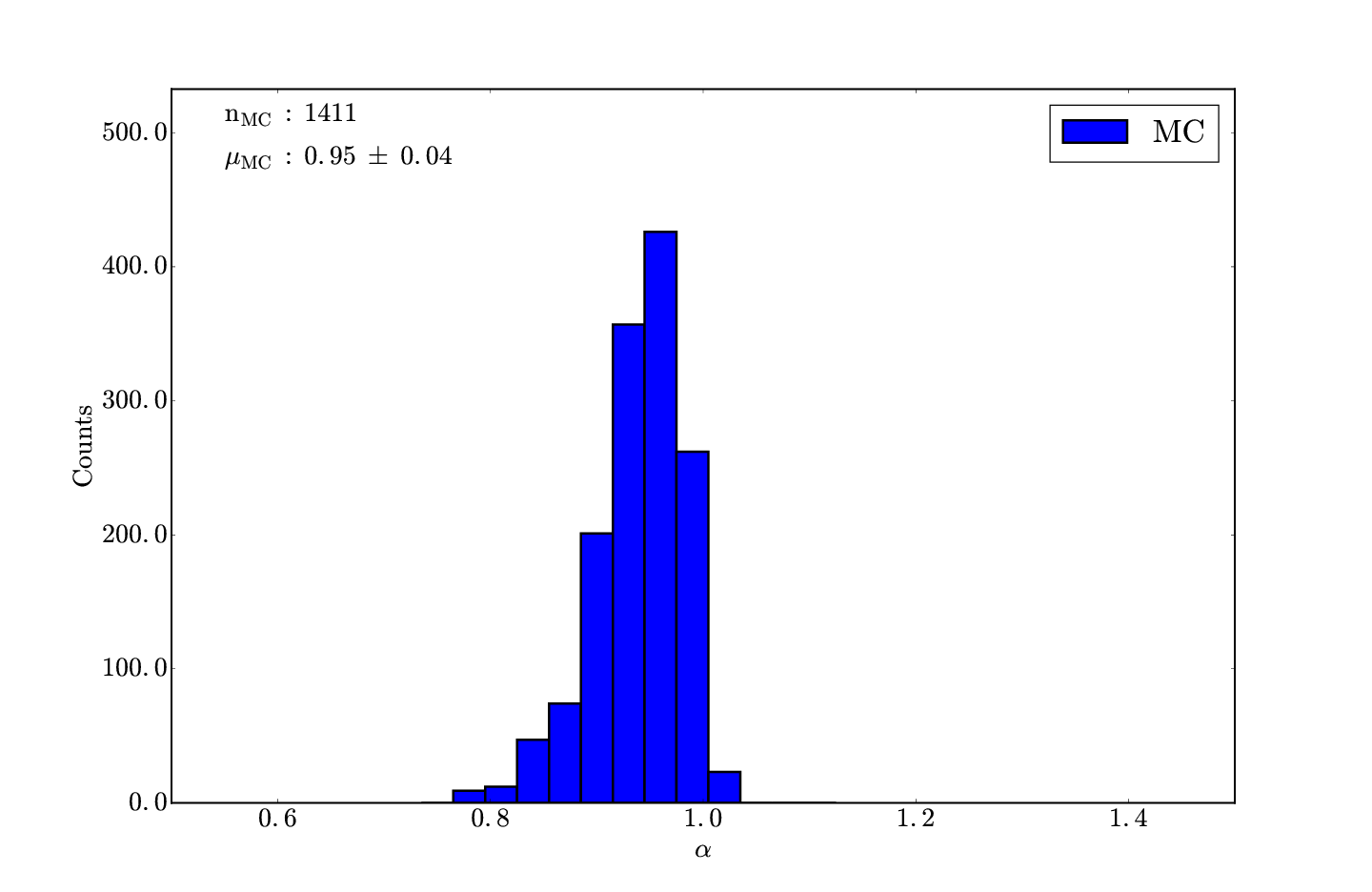}
\else
I am not enabling plots.
\fi
\caption{The same as Fig.~\ref{fig.KarinMC} with $\sim$1,400 trials repeating the V-shape technique for the Brangane family. The mean of the distribution is centered at $\alpha$ = 0.95 $\pm$ 0.04 and the bin size in the histogram is 0.03.}
\label{fig.BranganeMC}
\end{figure}

 $\sim$1,400 Monte Carlo runs were completed where $H$ magnitudes were randomised by the typical magnitude uncertainty of 0.25 for asteroids in the MPC catalogue \citep[][]{Oszkiewicz2011,Veres2015} and their $\pv$ was assumed to be 0.10 with an uncertainty of 0.03 \citep[][]{Masiero2013}. The value of $\alpha$ in 1,400 Monte Carlo trials is on average $\sim$0.95 with a RMS uncertainty of 0.04 as seen in Fig.~\ref{fig.BranganeMC}. The $\alpha$ distribution of Monte Carlo runs is slightly negatively skewed such that the most probable value is slightly higher than the mean of $0.95$. The value of $\mu_{\alpha} \; \lesssim \; $ 1.0 and a smaller value of $C\; = \; 3.90 \; \times 10^{-6}$ au compared to $C_{EV} \; = \; 7.24 \; \times 10^{-6}$ au calculated from Eq.~\ref{eqn.VEVvsCalphaFinal} assuming $V_{EV}$ = 23 $\mps$, the escape velocity from a parent body with a parent body diameter, $D_{pb} \; = \; 42$ km \citep[][]{Broz2013b} and $\rho \; = \; 2.3 \; \gpcmc$ typical for S-type asteroids \citep[][]{Carry2012} suggests that the spread of fragments in the Brangane family in $a$ vs. $D_r$ space is mostly due to the ejection velocity of the fragments with only moderate modification in $a$ due to the Yarkovsky effect.
 
\subsection{Brasilia}
\label{s.Brasiliasection}

\begin{figure}
\centering
\hspace*{-0.7cm}
\ifincludeplots
\includegraphics[scale=0.425]{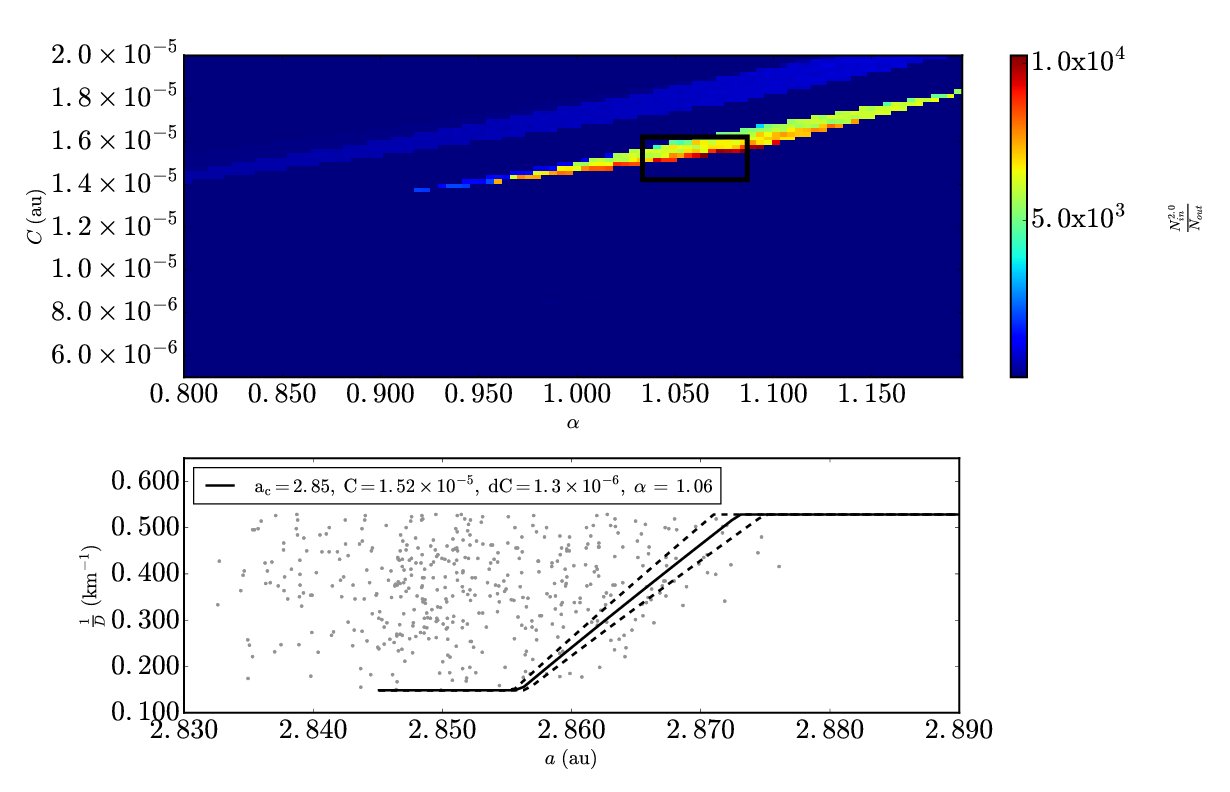}
\else
I am not enabling plots.
\fi
\caption{The same as Fig.~\ref{fig.synErig0Myrs} for Brasilia asteroid family data from \citet[][]{Nesvorny2015a}. (Top panel) $\Delta \alpha$ is equal to $4.0 \times 10^{-3}$ au and $\Delta C$, is equal to $6.5 \times 10^{-8}$ au. (Bottom Panel) $D_r(a,a_c,C\pm dC,\pv,\alpha)$ is plotted with $\pv = 0.24$, $a_c$ = 2.855 au and $dC \; = \; 1.3 \x 10^{-6}$ au.}
\label{fig.BrasiliaAlpha}
\end{figure}

The X-type fragment Brasilia asteroid family first identified by \citet[][]{Zappala1995} located in the outer Main Belt. The M/N solar system dust band was later attributed to the formation of the family with the asteroid 293 Brasilia being an interloper in its own family \citep[][]{Nesvorny2003, Broz2013b}. The asteroid 1521 Sejnajoki is more likely to be the largest asteroid family member in the Brasilia asteroid family, but we will use Brasilia as the name for the asteroid family. The age of the Brasilia family is estimated to be 50$\pm$40 Myrs by \citet[][]{Broz2013b}. The V-shape identification technique was applied to 584 asteroids belonging to the Brasilia asteroid family defined by \citet[][]{Nesvorny2015a}. Eqs.~\ref{eq.border_method_N_outer} and \ref{eq.border_method_N_inner} are integrated using the interval [$a_c$,$\infty$) for the Dirac delta function $\delta(a_{j}-a)$ because the inner half of the family V-shape for Brasilia is clipped due to the presence of the 5:2 MMR at 2.81 au as seen in the bottom panel of Fig.~\ref{fig.BrasiliaAlpha}. The interval [$0.13,0.53$] for the Dirac delta function $\delta(D_{r,j}-D_r )$ was used  to cover the majority of the range of the outer V-shape half while excluding possible interlopers at larger values of $D_r$.  Eq.~\ref{eqn.apvDvsCfinal} is truncated to 0.13 for $D_r$ $ <$ $0.13$ and to 0.53 for $D_r$ $>$ 0.53. Asteroid $H$ values were converted to $D$ using Eq.~\ref{eq.HtoD} using the value of $\pv$ = 0.24 typical for members of the Brasilia family \citep[][]{Masiero2013, Spoto2015}.
    
 The peak in $\frac{N_{in}^2}{N_{out}}$ at $(a_c, \; C, \; \alpha) \; = \; (2.85 \; \mathrm{au}, \; 1.52 \times 10^{-6} \; \mathrm{au}, \;  \sim1.06)$ as seen in the top panel of Fig.~\ref{fig.BrasiliaAlpha} is located in the range 2.83 au $< \; a \;<$ 2.89 au, 5 $\times \; 10^{-6}$ au $< \; C \;<$ 2.0 $\times \; 10^{-5}$ au and 0.8 $< \; \alpha \;<$ 1.2.  A $dC \; = \; 1.3 \; \times 10^{-6}$ au was used. The technique was repeated with the Brasilia family defined by \citet[][]{Spoto2015} resulting in similar results.
 
\begin{figure}
\centering
\hspace*{-1.1cm}
\ifincludeplots
\includegraphics[scale=0.3725]{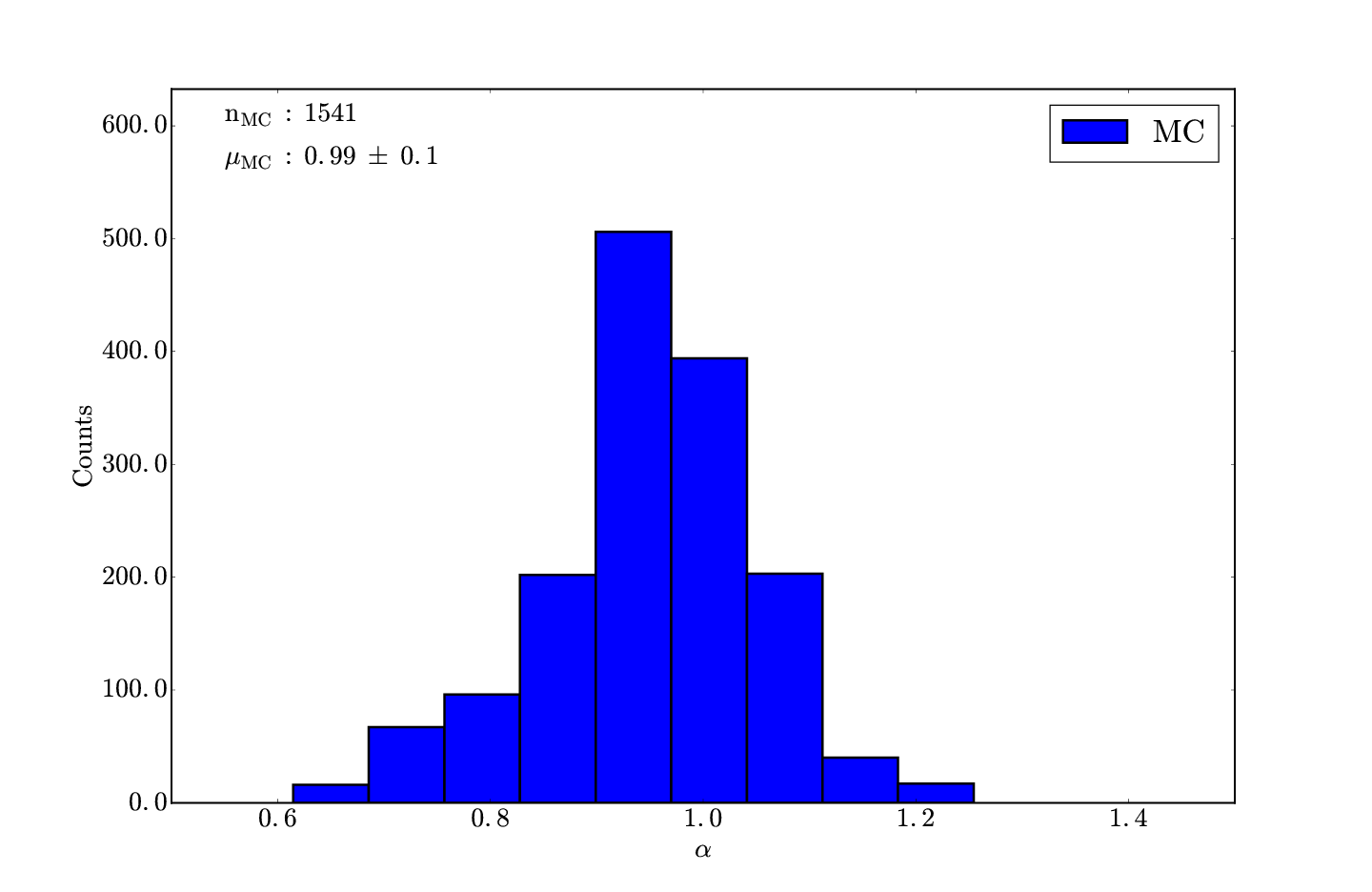}
\else
I am not enabling plots.
\fi
\caption{The same as Fig.~\ref{fig.KarinMC} with $\sim$1,500 trials repeating the V-shape technique for the Brasilia family. The mean of the distribution is centered at $\alpha$ $\simeq$ 1.0 $\pm$ 0.10 and the bin size in the histogram is 0.07.}
\label{fig.BrasiliaMC}
\end{figure}

 $\sim$1,500 Monte Carlo runs were completed where $H$ magnitudes were randomised by the typical magnitude uncertainty of 0.25 for asteroids in the MPC catalogue \citep[][]{Oszkiewicz2011,Veres2015} and their $\pv$ was assumed to be 0.24 with an uncertainty of 0.06 \citep[][]{Masiero2013}. The value of $\alpha$ in $\sim$1,500 Monte Carlo trials is on average $\sim$0.99 with a RMS uncertainty of 0.1 as seen in Fig.~\ref{fig.BrasiliaMC}. The value of $\mu_{\alpha} \; \sim \; $ 1.0 and a similar value of $C\; = \; 1.5 \; \times 10^{-5}$ au compared to $C_{EV} \; = \; 1.3 \; \times 10^{-5}$ au calculated from Eq.~\ref{eqn.VEVvsCalphaFinal} assuming $V_{EV}$ = 22 $\mps$, the escape velocity from a parent body with a parent body diameter, $D_{pb} \; = \; 34$ km \citep[][]{Broz2013b} and $\rho \; = \; 2.7 \; \gpcmc$ typical for X-type asteroids \citep[][]{Carry2012} suggests that the spread of fragments in the Brasilia family in $a$ vs. $D_r$ space is mostly due to the ejection velocity of the fragments with only moderate modification in $a$ due to the Yarkovsky effect.
 
\subsection{Clarissa}

\begin{figure}
\centering
\hspace*{-0.7cm}
\ifincludeplots
\includegraphics[scale=0.425]{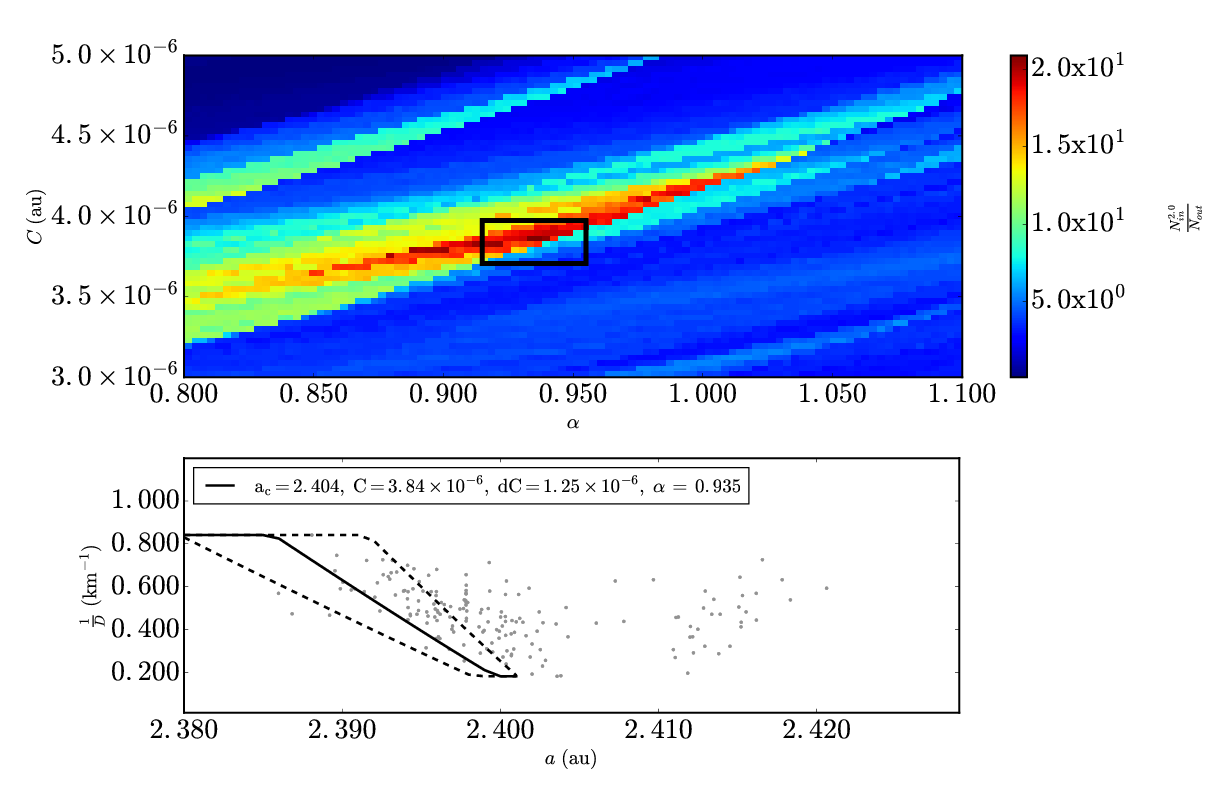}
\else
I am not enabling plots.
\fi
\caption{The same as Fig.~\ref{fig.synErig0Myrs} for Clarissa asteroid family data from \citet[][]{Nesvorny2015a}. (Top panel) $\Delta \alpha$ is equal to $3.0 \times 10^{-3}$ au and $\Delta C$, is equal to $3.5 \times 10^{-8}$ au. (Bottom Panel) $D_r(a,a_c,C\pm dC,\pv,\alpha)$ is plotted with $\pv = 0.06$, $a_c$ = 2.404 au and $dC \; = \; 1.25 \x 10^{-6}$ au.}
\label{fig.ClarissaAlph}
\end{figure}

The C complex Clarissa asteroid family located in the inner Main Belt was first identified by \citet[][]{Nesvorny2010a}. The age of the family is estimated to be $<100$ Myrs by \citet[][]{Broz2013b}. The V-shape identification technique was applied to 179 asteroids belonging to the Clarissa asteroid family defined by \citet[][]{Nesvorny2015a}. Eqs.~\ref{eq.border_method_N_outer} and \ref{eq.border_method_N_inner} are integrated using the interval ($-\infty$,$a_c$] for the Dirac delta function $\delta(a_{j}-a)$ because the outer half of the family V-shape for Clarissa has a fewer asteroids in $a$ vs, $D_r$ space on its outer V-shape half as seen panel of Fig.~\ref{fig.ClarissaAlph}. The interval [$0.11, 0.91$] for the Dirac delta function $\delta(D_{r,j}-D_r )$ was used to cover the full range of fragments in the Clarissa family inner half V-shape.  Eq.~\ref{eqn.apvDvsCfinal} is truncated to 0.11 for $D_r$ $ <$ $0.11$ and to 0.91 for $D_r$ $>$ 0.91. Asteroid $H$ values were converted to $D$ using Eq.~\ref{eq.HtoD} using the value of $\pv$ = 0.06 typical for members of the Clarissa family \citep[][]{Masiero2013}.

 The peak in $\frac{N_{in}^2}{N_{out}}$ at $(a_c, \; C, \; \alpha) \; = \; (2.404 \; \mathrm{au}, \; 3.84 \times 10^{-6} \; \mathrm{au}, \;  \sim0.94)$ as seen in the top panel of Fig.~\ref{fig.ClarissaAlph} is located in the range 2.38 au $< \; a \;<$ 2.43 au, 3.0 $\times \; 10^{-6}$ au $< \; C \;<$ 5.0 $\times \; 10^{-6}$ au and 0.8 $< \; \alpha \;<$ 1.1.  A $dC \; = \; 1.25 \; \times 10^{-6}$ au was used. 
 
\begin{figure}
\centering
\hspace*{-1.1cm}
\ifincludeplots
\includegraphics[scale=0.3725]{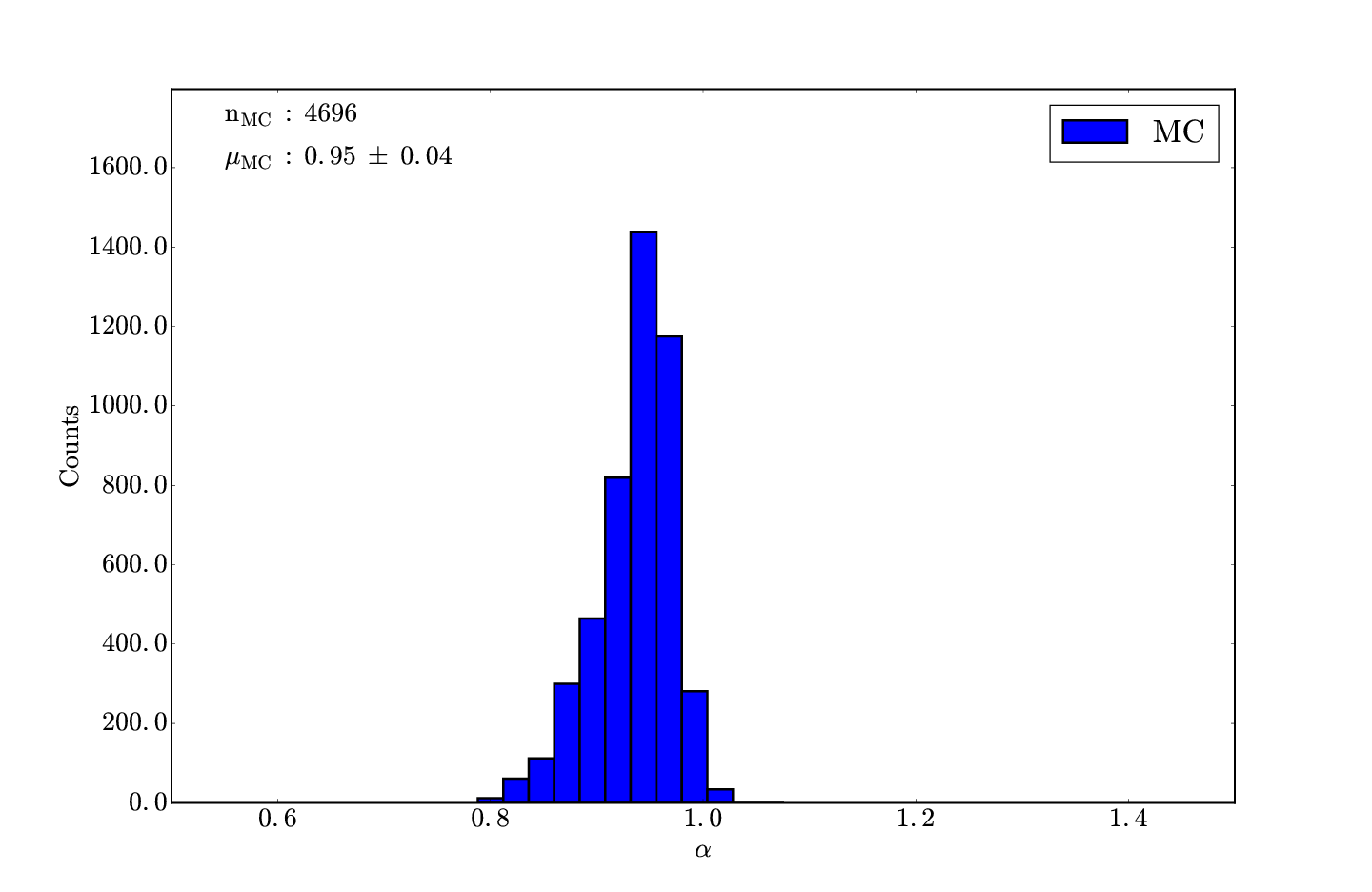}
\else
I am not enabling plots.
\fi
\caption{The same as Fig.~\ref{fig.KarinMC} with $\sim$4,700 trials repeating the V-shape technique for the Clarissa family. The mean of the distribution is centered at $\alpha$ = 0.95 $\pm$ 0.04 and the bin size in the histogram is 0.07.}
\label{fig.ClarissaMC}
\end{figure}

 $\sim$4,700 Monte Carlo runs were completed where $H$ magnitudes were randomised by the typical magnitude uncertainty of 0.25 for asteroids in the MPC catalogue \citep[][]{Oszkiewicz2011,Veres2015} and their $\pv$ was assumed to be 0.06 with an uncertainty of 0.02 \citep[][]{Masiero2013}. The value of $\alpha$ in $\sim$4,700 Monte Carlo trials is on average $\sim$0.95 with a RMS uncertainty of 0.04 as seen in Fig.~\ref{fig.ClarissaMC}. The $\alpha$ distribution of Monte Carlo runs is slightly negatively skewed such that the most probable value is slightly higher than the mean of $0.95$. The value of $\mu_{\alpha} \; \lesssim \; $ 1.0 and a similar value of $C\; = \; 3.84 \; \times 10^{-6}$ au compared to $C_{EV} \; = \; 3.91 \; \times 10^{-6}$ au calculated from Eq.~\ref{eqn.VEVvsCalphaFinal} assuming $V_{EV}$ = 17 $\mps$, the escape velocity from a parent body with a parent body diameter, $D_{pb} \; = \; 39$ km \citep[][]{Broz2013b} and $\rho \; = \; 1.4 \; \gpcmc$ typical for C-type asteroids \citep[][]{Carry2012} suggests that the spread of fragments in the Clarissa family in $a$ vs. $D_r$ space is mostly due to the ejection velocity of the fragments with only moderate modification in $a$ due to the Yarkovsky effect.
 
\subsection{Iannini}
\label{s.iannini}

The S-type Iannini asteroid family located in the central Main Belt and the presence of the J/K solar system dust band is attributed to the formation of the family \citep[][]{Nesvorny2003, Willman2008}. The age of the Iannini family is estimated to be 5$\pm$5 Myrs by \citet[][]{Broz2013b}. The V-shape identification technique was applied to 584 asteroids belonging to the Iannini asteroid family defined by \citet[][]{Nesvorny2015a}. Eqs.~\ref{eq.border_method_N_outer} and \ref{eq.border_method_N_inner} are integrated using the interval ($-\infty$,$\infty$) for the Dirac delta function $\delta(a_{j}-a)$ and the interval [$0.4,1.5$] is used for the Dirac delta function $\delta(D_{r,j}-D_r )$ to contain fragments defining the border of the Iannini family V-shape.  Eq.~\ref{eqn.apvDvsCfinal} is truncated to 0.4 for $D_r$ $ <$ $0.4$ and to 1.5 for $D_r$ $>$ 1.5. Asteroid $H$ values were converted to $D$ using Eq.~\ref{eq.HtoD} using the value of $\pv$ = 0.32 typical for members of the Iannini family \citep[][]{Masiero2013}. 

The technique was repeated with the Iannini family defined by \citet[][]{Spoto2015}. The Iannini family defined by \citet[][]{Spoto2015} differs from the definition of the Iannini family by \citet[][]{Nesvorny2015a} by using the asteroid Nele as the largest fragment in the family. If the 17 km diameter asteroid Nele is considered to be the largest remaining fragment of the Iannini family, the family V-shape is slightly asymmetric \citet[][]{Spoto2015}. Regardless of the potential asymmetric V-shape of the Iannini/Nele family, the V-shape has a symmetrical shape in both \citet[][]{Spoto2015} and \citet[][]{Nesvorny2015a} definitions of the Iannini family in the interval $0.4$ $\leq$ $D_r$ $\leq$ $1.5$. This interval does not include asteroids Iannini and Nele which have $D_r$ equal to 0.2 km$^{-1}$ and 0.06 km$^{-1}$ respectively, and the results of the $a_c$, $C$ and $\alpha$ determination technique are similar when applied to both catalogues. An alternative explanation to the membership of Nele to the Iannini family is that it is an interloper because of how far it is offset from the apex of the main family V-shape \citep[][]{Nesvorny2015a}.

\begin{figure}
\centering
\hspace*{-0.7cm}
\ifincludeplots
\includegraphics[scale=0.425]{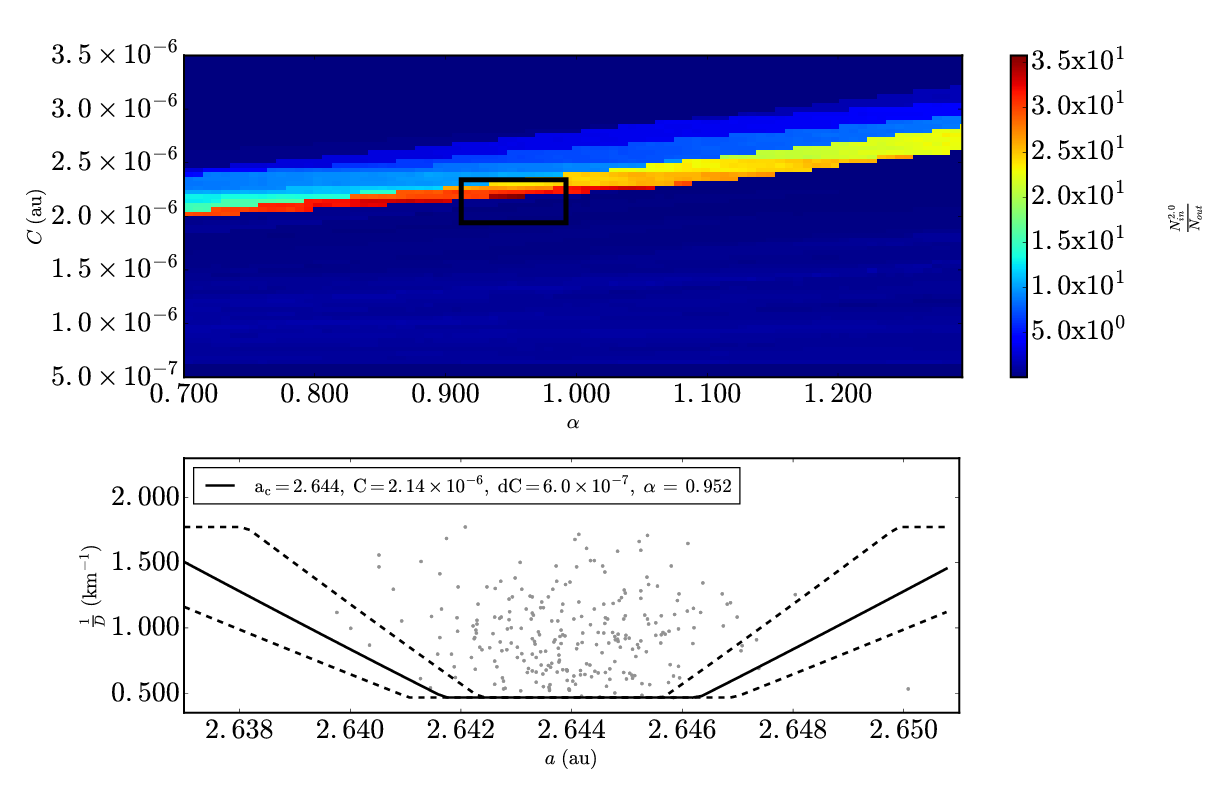}
\else
I am not enabling plots.
\fi
\caption{The same as Fig.~\ref{fig.synErig0Myrs} for Iannini asteroid family data from \citet[][]{Nesvorny2015a}. (Top panel) $\Delta \alpha$ is equal to $7.0 \times 10^{-3}$ au and $\Delta C$, is equal to $4.0 \times 10^{-8}$ au. (Bottom Panel) $D_r(a,a_c,C\pm dC,\pv,\alpha)$ is plotted with $\pv = 0.36$, $a_c$ = 2.644 au and $dC \; = \; 6.0 \x 10^{-7}$ au.}
\label{fig.IanniniAlph}
\end{figure}
    
 The peak in $\frac{N_{in}^2}{N_{out}}$ at $(a_c, \; C, \; \alpha) \; = \; (2.64 \; \mathrm{au}, \; 2.1 \times 10^{-6} \; \mathrm{au}, \;  \sim0.95)$ as seen in the top panel of Fig.~\ref{fig.IanniniAlph} is located in the range 2.637 au $< \; a \;<$ 2.65 au, 0.5 $\times \; 10^{-6}$ au $< \; C \;<$ 3.5 $\times \; 10^{-6}$ au and 0.7 $< \; \alpha \;<$ 1.3.  A $dC \; = \; 6.0 \; \times 10^{-7}$ au was used. The technique was repeated with the Iannini family defined by \citet[][]{Spoto2015} resulting in similar results.
 
\begin{figure}
\centering
\hspace*{-1.1cm}
\ifincludeplots
\includegraphics[scale=0.3725]{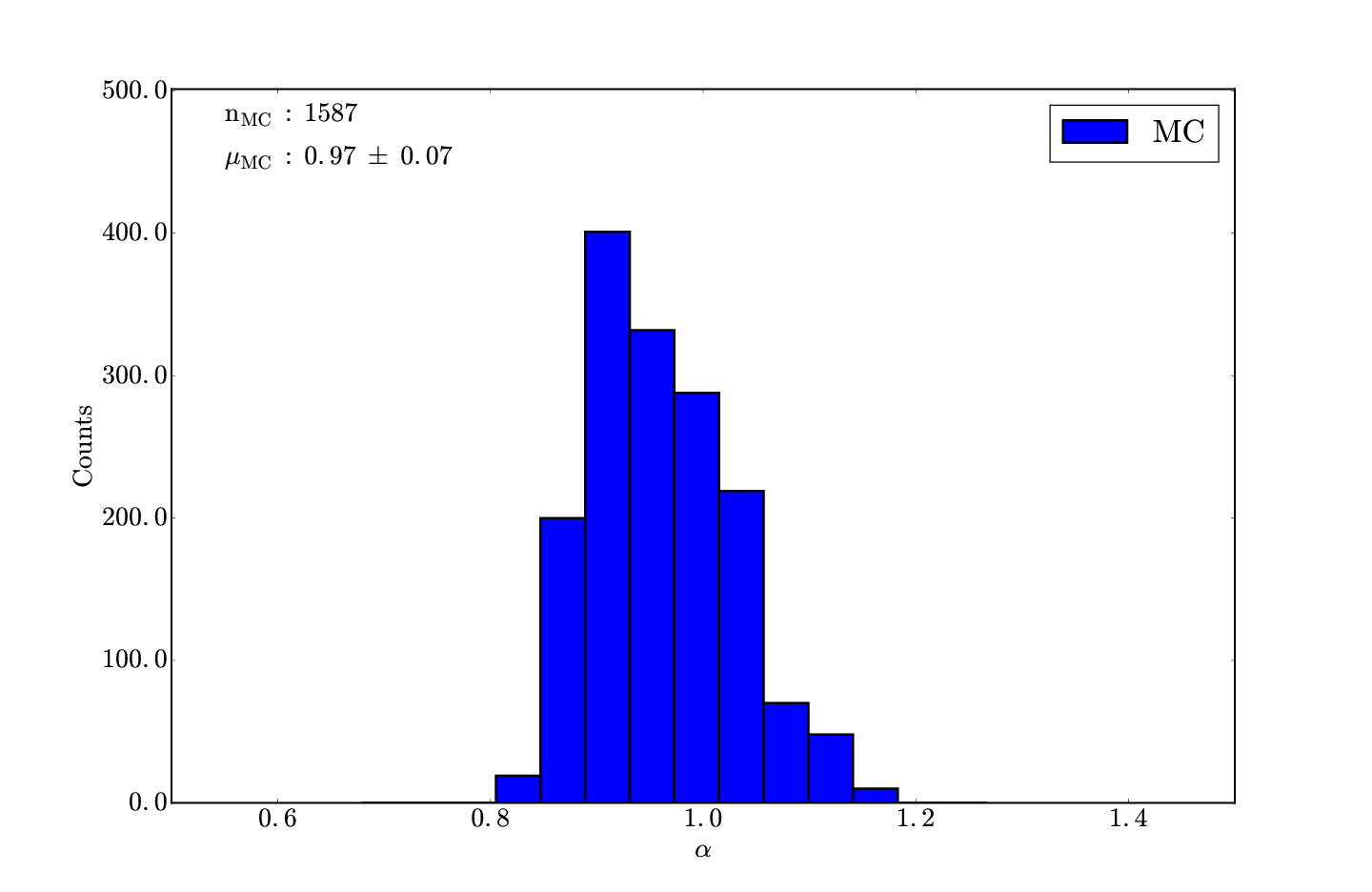}
\else
I am not enabling plots.
\fi
\caption{The same as Fig.~\ref{fig.KarinMC} with $\sim$1,600 trials repeating the V-shape technique for the Iannini family. The mean of the distribution is centered at $\alpha$ = 0.97 $\pm$ 0.07 and the bin size in the histogram is 0.042.}
\label{fig.IanniniMC}
\end{figure}

 $\sim$1,400 Monte Carlo runs were completed where $H$ magnitudes were randomised by the typical magnitude uncertainty of 0.25 for asteroids in the MPC catalogue \citep[][]{Oszkiewicz2011,Veres2015} and their $\pv$ was assumed to be 0.32 with an uncertainty of 0.1 \citep[][]{Masiero2013}. The value of $\alpha$ in $\sim$1,600 Monte Carlo trials ranges between 0.5 $\lesssim$ $\alpha$ $\lesssim$ 1.5  and is on average $\sim$0.97 with a RMS uncertainty of 0.07 as seen in Fig.~\ref{fig.IanniniMC}. The $\alpha$ distribution of Monte Carlo runs is slightly positively skewed such that the most probable value is slightly lower than the mean of $0.97$. The value of $\mu_{\alpha} \; \sim \; $ 1.0 and a smaller value of $C\; = \; 2.1 \; \times 10^{-6}$ au compared to $C_{EV} \; = \; 3.3 \; \times 10^{-6}$ au calculated from Eq.~\ref{eqn.VEVvsCalphaFinal} assuming $V_{EV}$ = $\sim$5.6 $\mps$, the escape velocity from a parent body with a parent body diameter, $D_{pb} \; = \; 10$ km \citep[][]{Nesvorny2015a} and $\rho \; = \; 2.3 \; \gpcmc$ typical for S-type asteroids \citep[][]{Carry2012} suggests that the spread of fragments in the Iannini family in $a$ vs. $D_r$ space is mostly due to the ejection velocity of the fragments with only moderate modification in $a$ due to the Yarkovsky effect. This conclusion is strengthened if Nele is considered to be the largest remnant asteroid of the Iannini family increasing $D_{pb}$ to 22 km which increases $C_{EV}$ to $7.3 \; \times 10^{-6}$ au, now more than three times larger than the measured value of $C\; = \; 2.1 \; \times 10^{-6}$ au for the family V-shape.

\subsection{Karin}
\label{s.karin}

The Karin asteroid family located in the outer Main Belt has an initial ejection velocity field that scales with $\alpha_{EV} \; = \; 1.0$ \citep[][]{Nesvorny2002b,Nesvorny2006a} and is known to contain S-type asteroids \citep[][]{Molnar2009, Harris2009}. The age of the Karin family is estimated to be 5.75$\pm$0.01 Myrs by \citet[][]{Carruba2016f}. The V-shape identification technique was applied to 429 asteroids belonging to the Karin asteroid family defined by \citet[][]{Nesvorny2015a}. Eqs.~\ref{eq.border_method_N_outer} and \ref{eq.border_method_N_inner} are integrated using the interval ($-\infty$,$\infty$) for the Dirac delta function $\delta(a_{j}-a)$ and the interval [$0.2,1.1$] is used for the Dirac delta function $\delta(D_{r,j}-D_r )$ to contain fragments on the border of the Karin family V-shape.  Eq.~\ref{eqn.apvDvsCfinal} is truncated to 0.2 for $D_r$ $ <$ $0.2$ and to 1.1 for $D_r$ $>$ 1.1 as seen in the bottom panel of Fig.~\ref{fig.KarinAlph}. Asteroid $H$ values were converted to $D$ using Eq.~\ref{eq.HtoD} using the value of $\pv$ = 0.21 typical for members of the Karin family \citep[][]{Harris2009}.
    
 The peak in $\frac{N_{in}^2}{N_{out}}$ at $(a_c, \; C, \; \alpha) \; = \; (2.86 \; \mathrm{au}, \; 3.1 \times 10^{-6} \; \mathrm{au}, \;  \sim1.05)$ as seen in the top panel of Fig.~\ref{fig.KarinAlph} is located in the range 2.850 au $< \; a \;<$ 2.875 au, 2.0 $\times \; 10^{-6}$ au $< \; C \;<$ 5.5 $\times \; 10^{-6}$ au and 0.7 $< \; \alpha \;<$ 1.3.  A $dC \; = \; 2.0 \; \times 10^{-6}$ au was used.
 
 $\sim$1,200 Monte Carlo runs were completed where $H$ magnitudes were randomised by the typical magnitude uncertainty of 0.25 for asteroids in the MPC catalogue \citep[][]{Oszkiewicz2011,Veres2015} and their $\pv$ was assumed to be 0.21 with an uncertainty of 0.06 \citep[][]{Harris2009}. The value of $\alpha$ in $\sim$1,200 Monte Carlo trials is on average $\sim$0.97 with a RMS uncertainty of 0.05 as seen in Fig.~\ref{fig.KarinMC}. The value of $\mu_{\alpha} \; \sim \; $ 1.0 and a smaller value of $C\; = \; 3.1 \; \times 10^{-6}$ au compared to $C_{EV} \; = \; 12.1 \; \times 10^{-6}$ au calculated from Eq.~\ref{eqn.VEVvsCalphaFinal} assuming $V_{EV}$ = $\sim$22.5 $\mps$, the escape velocity from a parent body with a parent body diameter, $D_{pb} \; = \; 40$ km \citep[][]{Durda2007} and $\rho \; = \; 2.3 \; \gpcmc$ typical for S-type asteroids \citep[][]{Carry2012} suggests that the spread of fragments in the Karin family in $a$ vs. $D_r$ space is mostly due to the ejection velocity of the fragments with only moderate modification in $a$ due to the Yarkovsky effect. This is in agreement with the results of \citet[][]{Carruba2016f} which found that the smallest fragments in the Karin family only drifted $\sim$10$^{-3}$ au over the lifetime of the family making only $\sim$10$^{-7}$ au of a difference in the value of the family V-shapes' $C$ value.
 
\subsection{K{\"o}nig}
\label{s.konig}

\begin{figure}
\centering
\hspace*{-0.7cm}
\ifincludeplots
\includegraphics[scale=0.425]{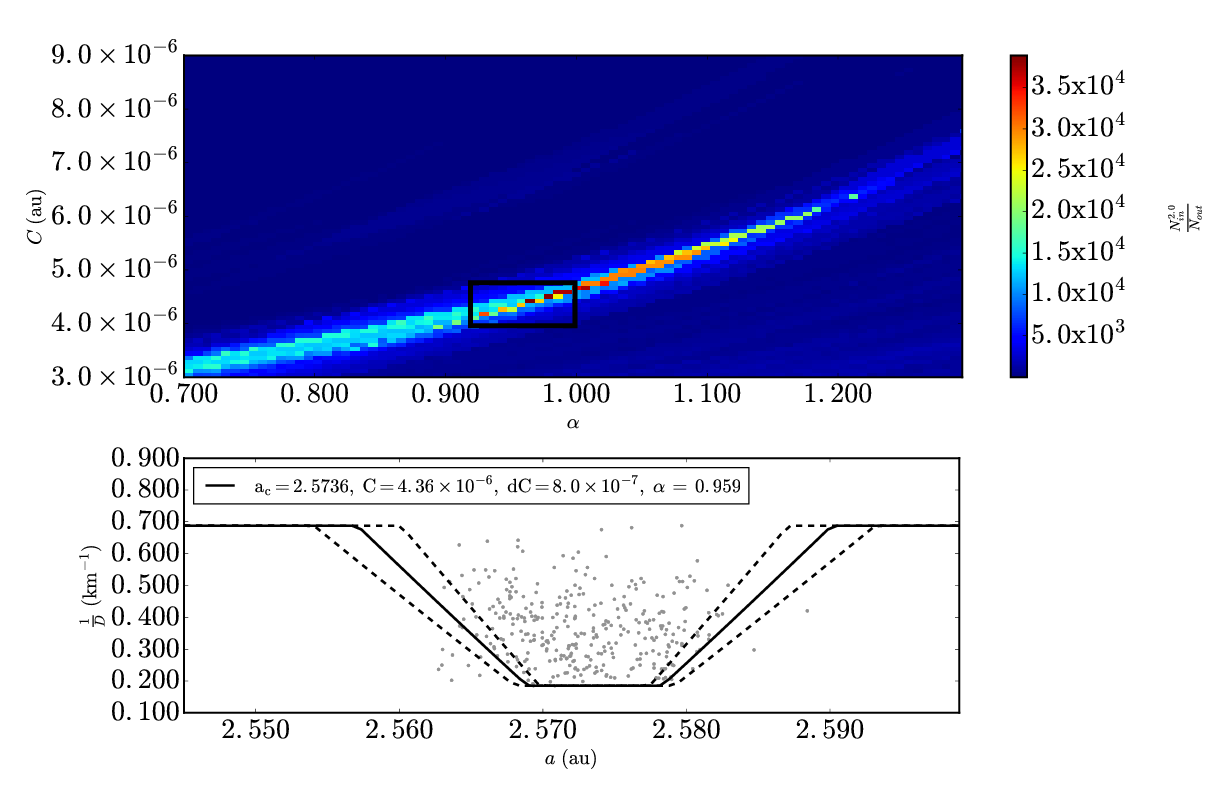}
\else
I am not enabling plots.
\fi
\caption{The same as Fig.~\ref{fig.synErig0Myrs} for K{\"o}nig asteroid family data from \citet[][]{Nesvorny2015a}. (Top panel) $\Delta \alpha$ is equal to $7.0 \times 10^{-3}$ au and $\Delta C$, is equal to $8.0 \times 10^{-8}$ au. (Bottom Panel) $D_r(a,a_c,C\pm dC,\pv,\alpha)$ is plotted with $\pv = 0.06$, $a_c$ = 2.574 au and $dC \; = \; 8.0 \x 10^{-7}$ au.}
\label{fig.KonigBorderAlph}
\end{figure}

The K{\"o}nig asteroid family located in the central Main Belt its formation is attributed to the G/H solar system dust band \citep[][]{Nesvorny2003} and is known to contain C-type asteroids \citep[][]{Broz2013b}. The age of the K{\"o}nig family is estimated to be between 50 - 100 Myrs \citep[][]{Spoto2015}. The V-shape identification technique was applied to 315 asteroids belonging to the K{\"o}nig asteroid family defined by \citet[][]{Nesvorny2015a}. Eqs.~\ref{eq.border_method_N_outer} and \ref{eq.border_method_N_inner} are integrated using the interval ($-\infty$,$\infty$) for the Dirac delta function $\delta(a_{j}-a)$ and the interval [$0.18,0.7$] is used for the Dirac delta function $\delta(D_{r,j}-D_r )$ to contain fragments  on the border of the K{\"o}nig family V-shape as seen in the bottom panel of Fig.~\ref{fig.KonigBorderAlph}.  Eq.~\ref{eqn.apvDvsCfinal} is truncated to 0.18 for $D_r$ $ <$ $0.18$ and to 0.7 for $D_r$ $>$ 0.7. Asteroid $H$ values were converted to $D$ using Eq.~\ref{eq.HtoD} using the value of $\pv$ = 0.06 typical for members of the K{\"o}nig family \citep[][]{Spoto2015}.

 The peak in $\frac{N_{in}^2}{N_{out}}$ at $(a_c, \; C, \; \alpha) \; = \; (2.57 \; \mathrm{au}, \; 4.4 \times 10^{-6} \; \mathrm{au}, \;  \sim0.96)$ as seen in the top panel of Fig.~\ref{fig.KonigBorderAlph} is located in the range 2.55 au $< \; a \;<$ 2.60 au, 3.0 $\times \; 10^{-6}$ au $< \; C \;<$ 9.0 $\times \; 10^{-6}$ au and 0.7 $< \; \alpha \;<$ 1.3.  A $dC \; = \; 8.0 \; \times 10^{-7}$ au was used. The technique was repeated with the K{\"o}nig family defined by \citet[][]{Spoto2015} resulting in similar results.
 
\begin{figure}
\centering
\hspace*{-1.1cm}
\ifincludeplots
\includegraphics[scale=0.3725]{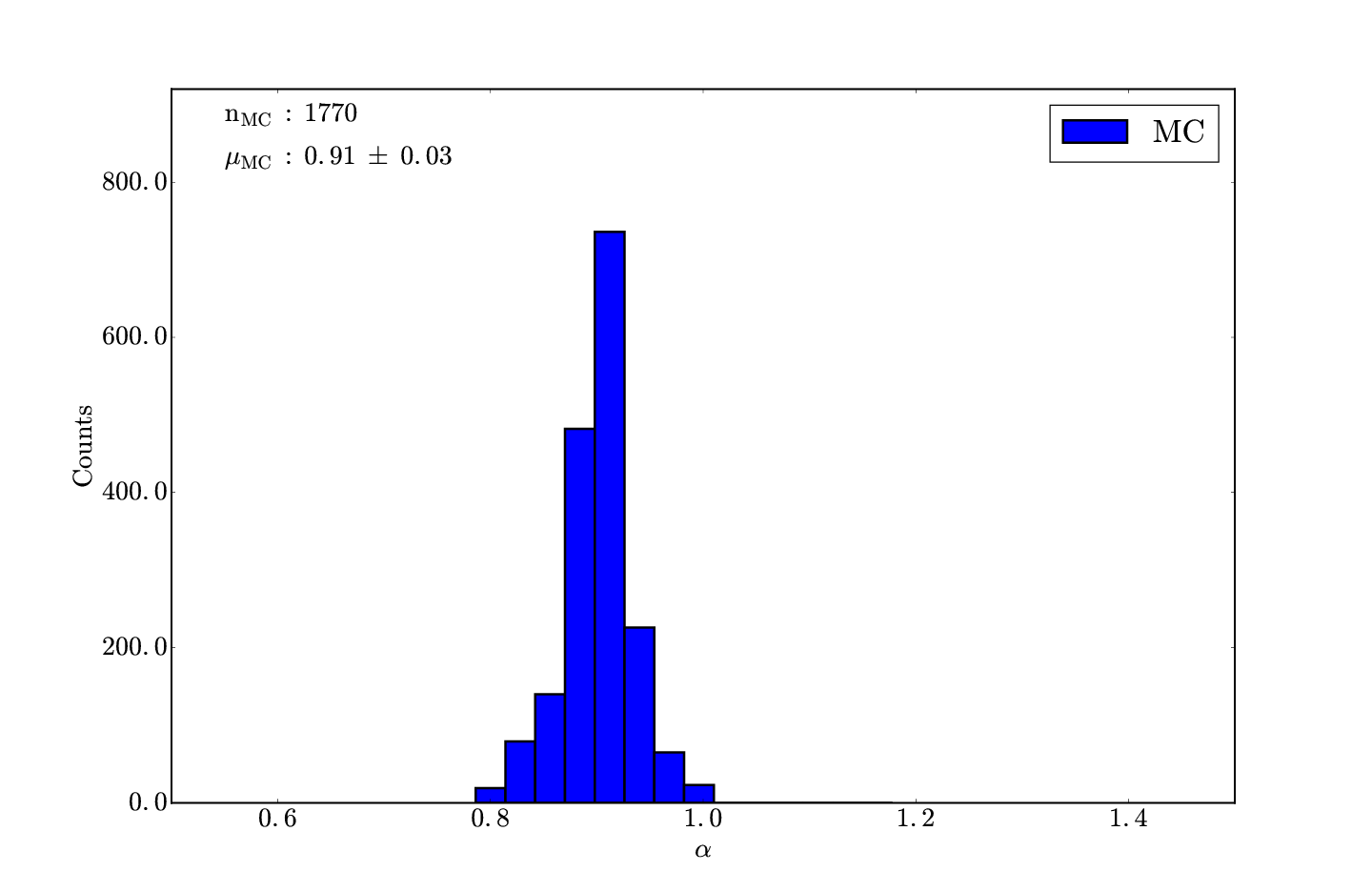}
\else
I am not enabling plots.
\fi
\caption{The same as Fig.~\ref{fig.KarinMC} with $\sim$1,700 trials repeating the V-shape technique for the K{\"o}nig family. The mean of the distribution is centered at $\alpha$ = 0.92 $\pm$ 0.03 and the bin size in the histogram is 0.03.}
\label{fig.konigMC}
\end{figure}

 $\sim$1,600 Monte Carlo runs were completed where $H$ magnitudes were randomised by the typical magnitude uncertainty of 0.25 for asteroids in the MPC catalogue \citep[][]{Oszkiewicz2011,Veres2015} and their $\pv$ was assumed to be 0.06 with an uncertainty of 0.02 \citep[][]{Spoto2015}. The value of $\alpha$ in $\sim$1,600 Monte Carlo trials is on average $\sim$0.92 with a RMS uncertainty of 0.03 as seen in Fig.~\ref{fig.konigMC}. The value of $\mu_{\alpha} \; \sim \; $ 0.92 and a similar value of $C\; = \; 4.4 \; \times 10^{-6}$ au compared to $C_{EV} \; = \; 3.3 \; \times 10^{-6}$ au calculated from Eq.~\ref{eqn.VEVvsCalphaFinal} assuming $V_{EV}$ = $\sim$14.6 $\mps$, the escape velocity from a parent body with a parent body diameter, $D_{pb} \; = \; 33$ km \citep[][]{Broz2013b} and $\rho \; = \; 1.4 \; \gpcmc$ typical for C-type asteroids \citep[][]{Carry2012} suggests that the spread of fragments in the K{\"o}nig family in $a$ vs. $D_r$ space is mostly due to the ejection velocity of the fragments with only moderate modification in $a$ due to the Yarkovsky effect.
 
\subsection{Koronis(2)}
\label{s.koronis2}

\begin{figure}
\centering
\hspace*{-0.7cm}
\ifincludeplots
\includegraphics[scale=0.425]{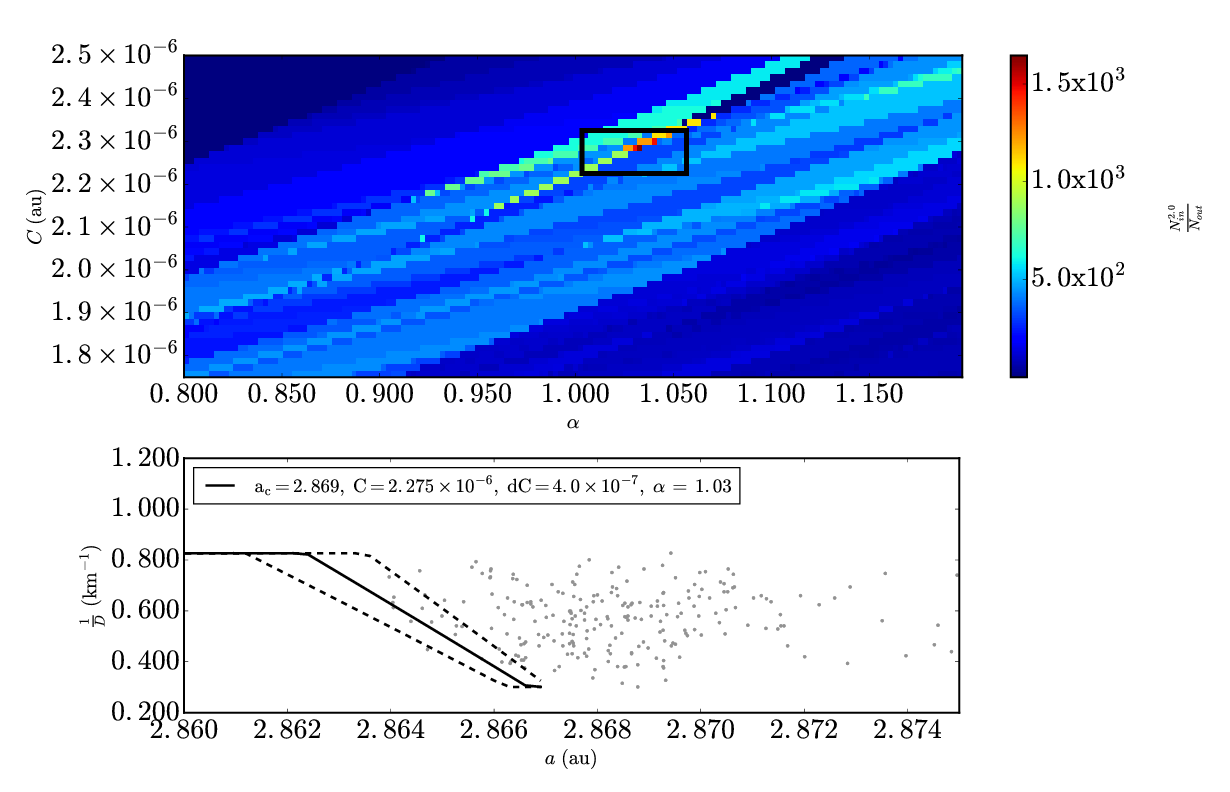}
\else
I am not enabling plots.
\fi
\caption{The same as Fig.~\ref{fig.synErig0Myrs} for Koronis(2) asteroid family data from \citet[][]{Nesvorny2015a}. (Top panel) $\Delta \alpha$ is equal to $2.5 \times 10^{-3}$ au and $\Delta C$, is equal to $1.5 \times 10^{-8}$ au. (Bottom Panel) $D_r(a,a_c,C\pm dC,\pv,\alpha)$ is plotted with $\pv = 0.14$, $a_c$ = 2.87 au and $dC \; = \; 4.0 \x 10^{-7}$ au.}
\label{fig.Koronis2Alph}
\end{figure}

The S-type Koronis(2) asteroid family located in the outer Main Belt was first identified by \citet[][]{Molnar2009}. The age of the family is estimated to be only 15 $\pm$ 5 Myrs by \citet[][]{Molnar2009} and \citet[][]{Broz2013b}. The V-shape identification technique was applied to 235 asteroids belonging to the Koronis(2) asteroid family defined by \citet[][]{Nesvorny2015a}. Eqs.~\ref{eq.border_method_N_outer} and \ref{eq.border_method_N_inner} are integrated using the interval ($-\infty$,$a_c$] for the Dirac delta function $\delta(a_{j}-a)$ because of the location of possible interlopers in the outer half of the family V-shape for Koronis(2) (see bottom panel of Fig.~\ref{fig.Koronis2Alph}). The interval [$0.28,0.89$] for the Dirac delta function $\delta(D_{r,j}-D_r )$ was used to mitigate the presence of potential interlopers in the inner half of the Koronis(2) family V-shape during the application of the V-shape identification technique.  Eq.~\ref{eqn.apvDvsCfinal} is truncated to 0.28 for $D_r$ $ <$ $0.28$ and to 0.89 for $D_r$ $>$ 0.89. Asteroid $H$ values were converted to $D$ using Eq.~\ref{eq.HtoD} using the value of $\pv$ = 0.14 typical for members of the Koronis(2) family \citep[][]{Masiero2013}.

 The peak in $\frac{N_{in}^2}{N_{out}}$ at $(a_c, \; C, \; \alpha) \; = \; (2.869 \; \mathrm{au}, \; 2.3 \times 10^{-6} \; \mathrm{au}, \;  \sim1.06)$ as seen in the top panel of Fig.~\ref{fig.Koronis2Alph} is located in the range 2.86 au $< \; a \;<$ 2.88 au, 1.8 $\times \; 10^{-6}$ au $< \; C \;<$ 2.5 $\times \; 10^{-6}$ au and 0.8 $< \; \alpha \;<$ 1.2.  A $dC \; = \; 4.0 \; \times 10^{-7}$ au was used. 
  
\begin{figure}
\centering
\hspace*{-1.1cm}
\ifincludeplots
\includegraphics[scale=0.3725]{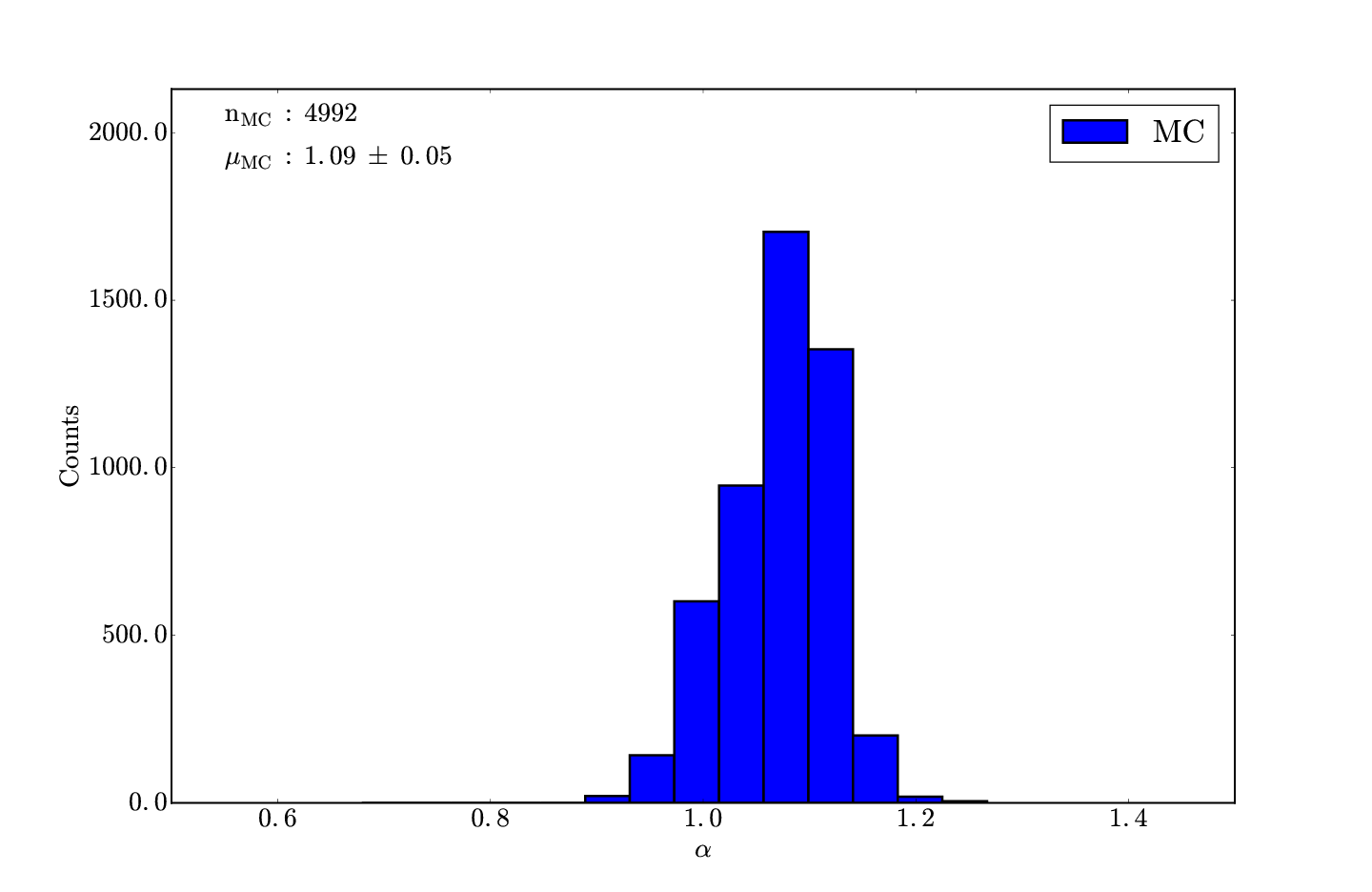}
\else
I am not enabling plots.
\fi
\caption{The same as Fig.~\ref{fig.KarinMC} with $\sim$5,000 trials repeating the V-shape technique for the Koronis(2) family. The mean of the distribution is centered at $\alpha$ = 1.09 $\pm$ 0.05 and the bin size in the histogram is 0.04.}
\label{fig.Koronis2MC}
\end{figure}

 $\sim$5,000 Monte Carlo runs were completed where $H$ magnitudes were randomised by the typical magnitude uncertainty of 0.25 for asteroids in the MPC catalogue \citep[][]{Oszkiewicz2011,Veres2015} and their $\pv$ was assumed to be 0.14 with an uncertainty of 0.04 \citep[][]{Spoto2015}. The value of $\alpha$ in $\sim$5,000 Monte Carlo trials is on average $\sim$1.1 with a RMS uncertainty of 0.05 as seen in Fig.~\ref{fig.Koronis2MC}. The $\alpha$ distribution of Monte Carlo runs is slightly negatively skewed such that the most probable value is slightly higher than the mean of $\sim1.1$. The value of $\mu_{\alpha} \; \simeq \; $ 1.1 and a smaller value of $C\; = \; 2.3 \; \times 10^{-6}$ au compared to $C_{EV} \; = \; 1.1 \; \times 10^{-5}$ au calculated from Eq.~\ref{eqn.VEVvsCalphaFinal} assuming $V_{EV}$ $\simeq$ 20 $\mps$, the escape velocity from a parent body with a parent body diameter, $D_{pb} \; = \; 35$ km \citep[][]{Nesvorny2015a} and $\rho \; = \; 2.3 \; \gpcmc$ typical for S-type asteroids \citep[][]{Carry2012} suggests that the spread of fragments in the Koronis(2) family in $a$ vs. $D_r$ space is due almost entirely to the ejection velocity of the fragments.
 
\subsection{Theobalda}
\label{s.theobalda}

\begin{figure}
\centering
\hspace*{-0.7cm}
\ifincludeplots
\includegraphics[scale=0.425]{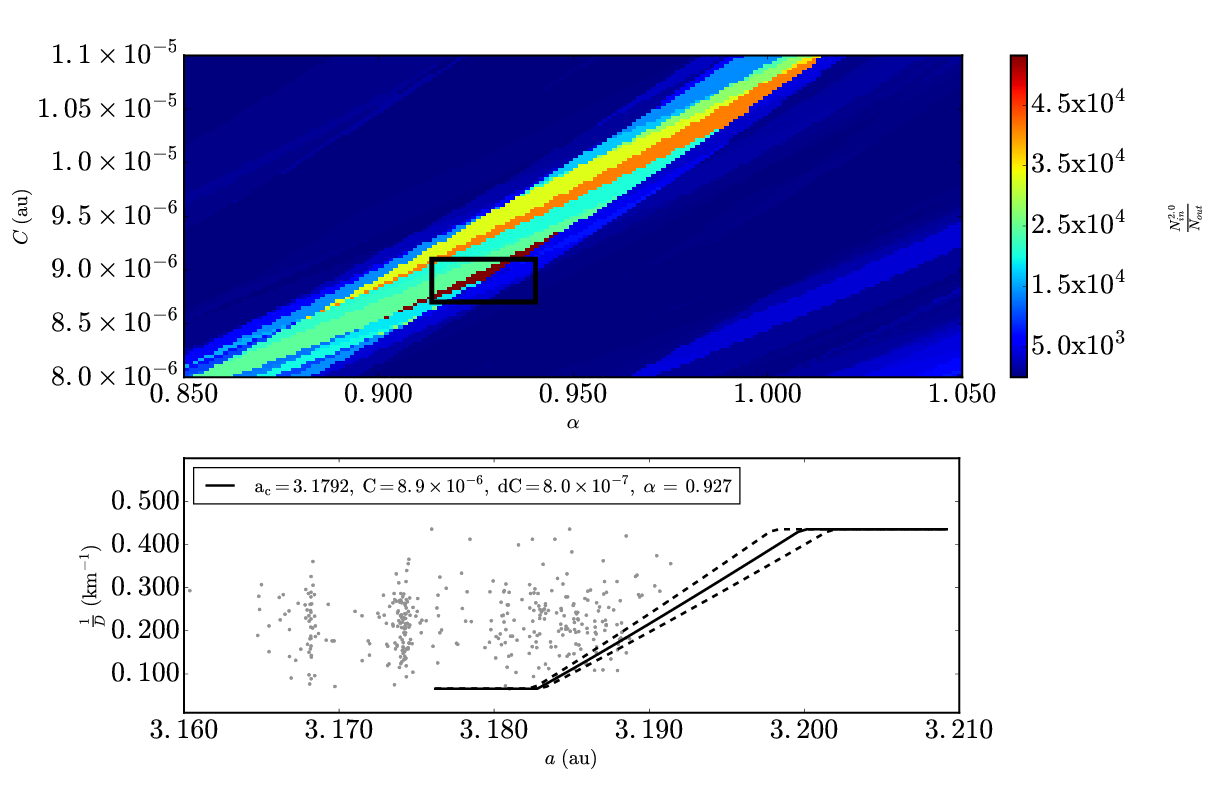}
\else
I am not enabling plots.
\fi
\caption{The same as Fig.~\ref{fig.synErig0Myrs} for Theobalda asteroid family data from \citet[][]{Nesvorny2015a}. (Top panel) $\Delta \alpha$ is equal to $1.3 \times 10^{-3}$ au and $\Delta C$, is equal to $3.0 \times 10^{-8}$ au. (Bottom Panel) $D_r(a,a_c,C\pm dC,\pv,\alpha)$ is plotted with $\pv = 0.06$, $a_c$ = 3.18 au and $dC \; = \; 8.0 \x 10^{-7}$ au.}
\label{fig.TheobaldaAlph}
\end{figure}

The 7 $\pm$ 2 Myrs old C-type Theobalda asteroid family located in the outer Main Belt was first identified by \citet[][]{Novakovic2010}.The V-shape identification technique was applied to 97 asteroids belonging to the Theobalda asteroid family defined by \citet[][]{Nesvorny2015a}. Eqs.~\ref{eq.border_method_N_outer} and \ref{eq.border_method_N_inner} are integrated using the interval [$a_c$,$\infty$) for the Dirac delta function $\delta(a_{j}-a)$ because the inner half of the family V-shape for Theobalda is affected by several secular resonances as seen Fig.~\ref{fig.TheobaldaAlph} \citep[][]{Novakovic2010}. The interval [$0.06,0.47$] for the Dirac delta function $\delta(D_{r,j}-D_r )$ was used to include asteroids on the outer V-shape border of the Theobalda family.  Eq.~\ref{eqn.apvDvsCfinal} is truncated to 0.06 for $D_r$ $ <$ $0.06$ and to 0.44 for $D_r$ $>$ 0.44. Asteroid $H$ values were converted to $D$ using Eq.~\ref{eq.HtoD} using the value of $\pv$ = 0.06 typical for members of the Theobalda family \citep[][]{Masiero2013}.

 The peak in $\frac{N_{in}^2}{N_{out}}$ at $(a_c, \; C, \; \alpha) \; = \; (3.18 \; \mathrm{au}, \; 8.9 \times 10^{-6} \; \mathrm{au}, \;  \sim0.93)$ as seen in the top panel of Fig.~\ref{fig.TheobaldaAlph} is located in the range 3.16 au $< \; a \;<$ 3.21 au, 8.0 $\times \; 10^{-6}$ au $< \; C \;<$ 1.2 $\times \; 10^{-5}$ au and 0.85 $< \; \alpha \;<$ 1.05.  A $dC \; = \; 8.0 \; \times 10^{-7}$ au was used. 
  
\begin{figure}
\centering
\hspace*{-1.1cm}
\ifincludeplots
\includegraphics[scale=0.3725]{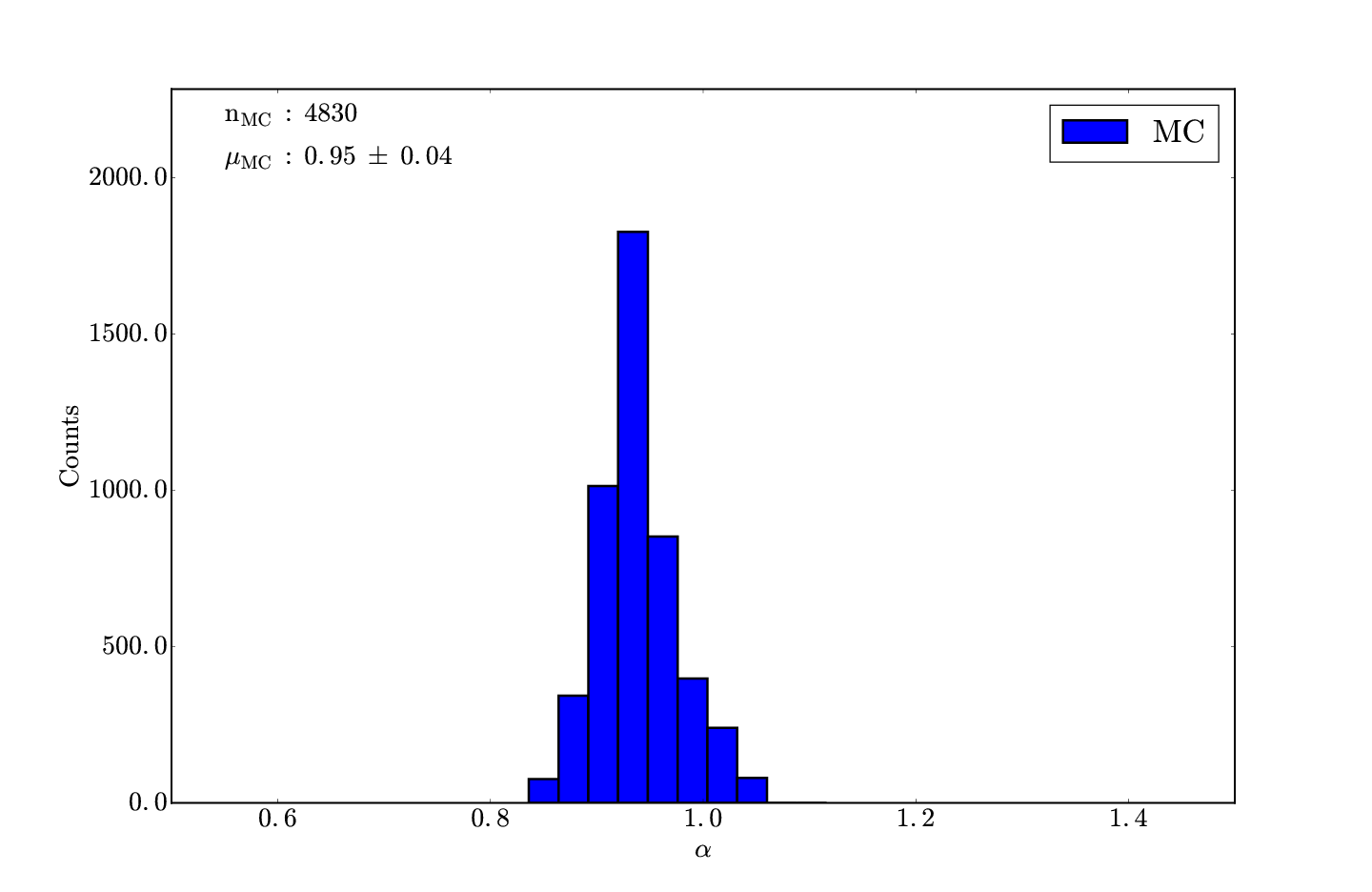}
\else
I am not enabling plots.
\fi
\caption{The same as Fig.~\ref{fig.KarinMC} with $\sim$4,800 trials repeating the V-shape technique for the Theobalda family. The mean of the distribution is centered at $\alpha$ = 0.95 $\pm$ 0.04 and the bin size in the histogram is 0.03.}
\label{fig.TheobaldaMC}
\end{figure}

4,830 Monte Carlo runs were completed where $H$ magnitudes were randomised by the typical magnitude uncertainty of 0.25 for asteroids in the MPC catalogue \citep[][]{Oszkiewicz2011,Veres2015} and their $\pv$ was assumed to be 0.06 with an uncertainty of 0.02 \citep[][]{Spoto2015}. The value of $\alpha$ in 5,200 Monte Carlo trials is on average $\sim$0.95 with a RMS uncertainty of 0.04 as seen in Fig.~\ref{fig.TheobaldaMC}. The value of $\mu_{\alpha} \; \simeq \; $ 0.95 and a smaller value of $C\; = \; 8.9 \; \times 10^{-6}$ au compared to $C_{EV} \; = \; 1.1 \; \times 10^{-5}$ au calculated from Eq.~\ref{eqn.VEVvsCalphaFinal} assuming $V_{EV}$ $\simeq$ 20 $\mps$, the escape velocity from a parent body with a parent body diameter, $D_{pb} \; = \; 74$ km \citep[][]{Nesvorny2015a} and $\rho \; = \; 1.4 \; \gpcmc$ typical for C-type asteroids \citep[][]{Carry2012} suggests that the spread of fragments in the Theobalda family in $a$ vs. $D_r$ space is due almost entirely to the ejection velocity of the fragments.
 
\subsection{Veritas}
\label{s.veritas}

\begin{figure}
\centering
\hspace*{-0.7cm}
\ifincludeplots
\includegraphics[scale=0.425]{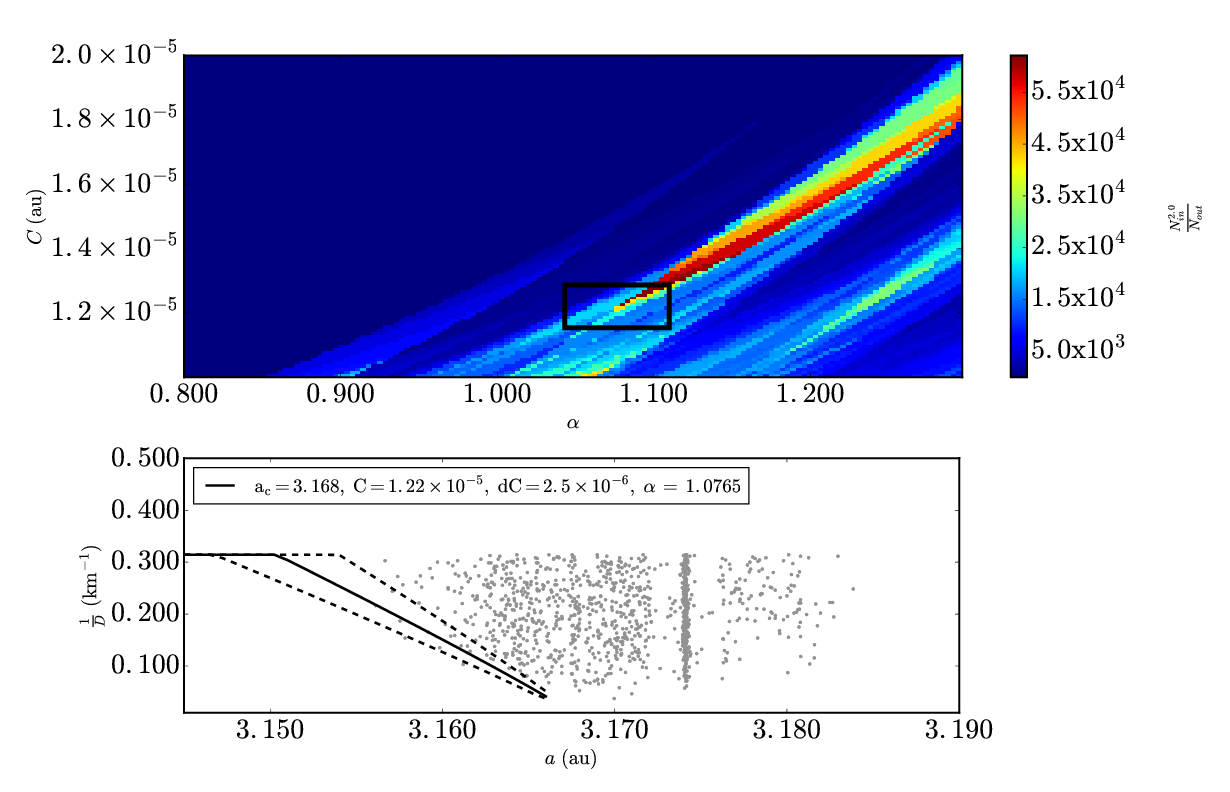}
\else
I am not enabling plots.
\fi
\caption{The same as Fig.~\ref{fig.synErig0Myrs} for Karin asteroid family data from \citet[][]{Nesvorny2015a}. (Top panel) $\Delta \alpha$ is equal to $3.5 \times 10^{-3}$ au and $\Delta C$, is equal to $1.0 \times 10^{-7}$ au. (Bottom Panel) $D_r(a,a_c,C\pm dC,\pv,\alpha)$ is plotted with $\pv = 0.07$, $a_c$ = 3.17 au and $dC \; = \; 2.5 \x 10^{-6}$ au.}
\label{fig.VeritasAlph}
\end{figure}

The Veritas family located in the outer MB and its formation is attributed to the $\gamma$ Infrared Astronomical Satellite dust band \citep[][]{Nesvorny2003}.The C-type family was first identified by \citet[][]{Zappala1990} and early studies proposed its age to be $<100$ Myrs old \citep[][]{Milani1994, Knezevic2002a}. Subsequent studies further constrained its age to 8.7 $\pm$ 1.7 Myrs \citep[][]{,Nesvorny2003,Tsiganis2007, Carruba2017a}. The V-shape identification technique was applied to 1135 asteroids belonging to the Veritas asteroid family defined by \citet[][]{Nesvorny2015a}. Eqs.~\ref{eq.border_method_N_outer} and \ref{eq.border_method_N_inner} are integrated using the interval ($-\infty$,$a_c$] for the Dirac delta function $\delta(a_{j}-a)$ because the outer half and centere of the family V-shape for Veritas is affected by several secular resonances as seen Fig.~\ref{fig.VeritasAlph} \citep[][]{Novakovic2010}. The interval [$0.03,0.51$] for the Dirac delta function $\delta(D_{r,j}-D_r )$ was used to remove interlopers in the inner V-shape half of the Veritas family during the application of the V-shape identification technique.  Eq.~\ref{eqn.apvDvsCfinal} is truncated to 0.03 for $D_r$ $ <$ $0.03$ and to 0.51 for $D_r$ $>$ 0.51. Asteroid $H$ values were converted to $D$ using Eq.~\ref{eq.HtoD} using the value of $\pv$ = 0.07 typical for members of the Veritas family \citep[][]{Masiero2013}.

 The peak in $\frac{N_{in}^2}{N_{out}}$ at $(a_c, \; C, \; \alpha) \; = \; (3.17 \; \mathrm{au}, \; 1.22 \times 10^{-5} \; \mathrm{au}, \;  1.08)$ as seen in the top panel of Fig.~\ref{fig.VeritasAlph} is located in the range 3.145 au $< \; a \;<$ 3.190 au, 1.0 $\times \; 10^{-5}$ au $< \; C \;<$ 2.0 $\times \; 10^{-5}$ au and 0.80 $< \; \alpha \;<$ 1.3.  A $dC \; = \; 2.5 \; \times 10^{-6}$ au was used.
  
\begin{figure}
\centering
\hspace*{-1.1cm}
\ifincludeplots
\includegraphics[scale=0.3725]{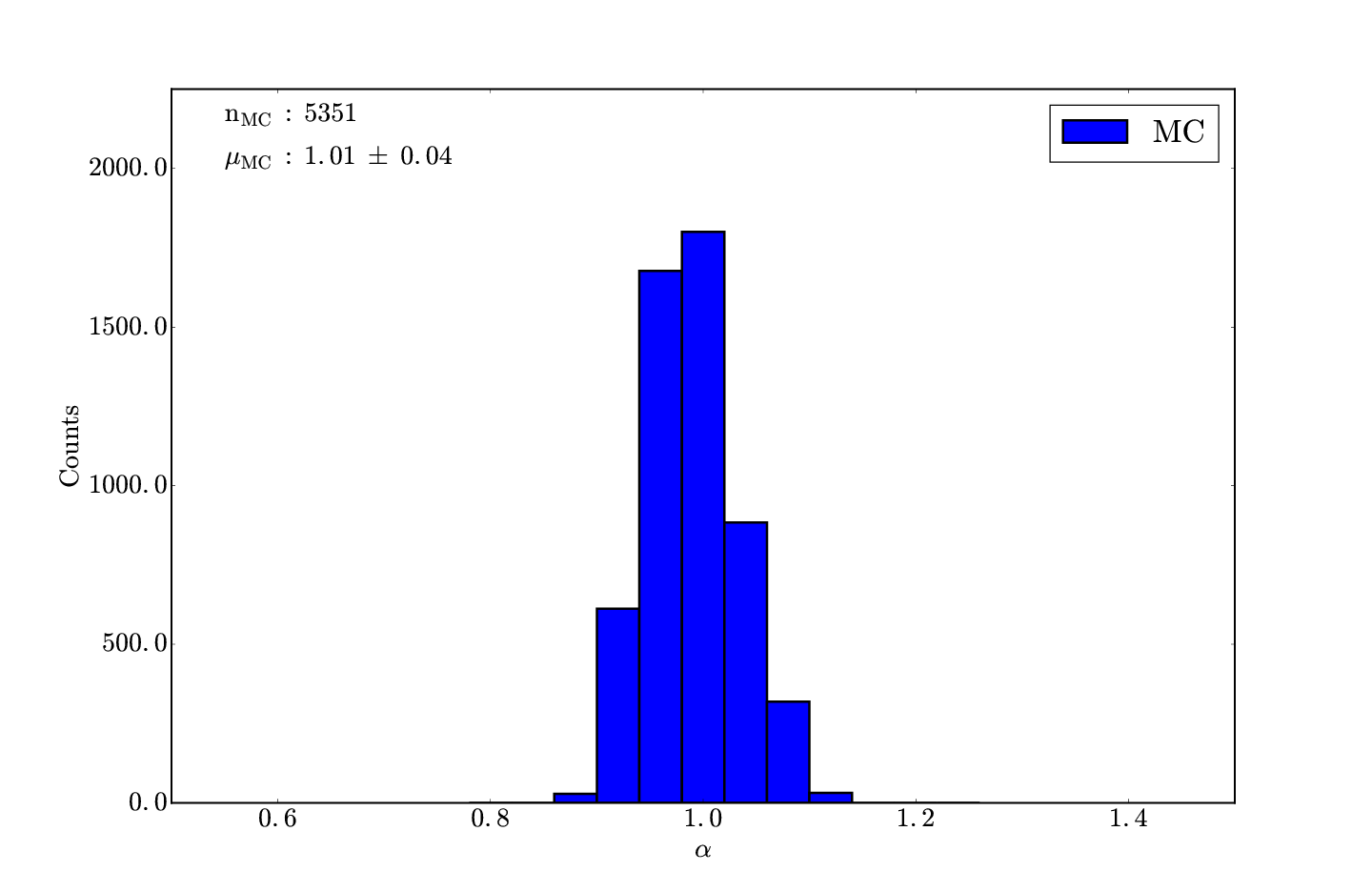}
\else
I am not enabling plots.
\fi
\caption{The same as Fig.~\ref{fig.KarinMC} with $\sim$5,400 trials repeating the V-shape technique for the Veritas family. The mean of the distribution is centered at $\alpha$ = 1.01 $\pm$ 0.04 and the bin size in the histogram is 0.04.}
\label{fig.VeritasMC}
\end{figure}

 $\sim$5,350 Monte Carlo runs were completed where $H$ magnitudes were randomised by the typical magnitude uncertainty of 0.25 for asteroids in the MPC catalogue \citep[][]{Oszkiewicz2011,Veres2015} and their $\pv$ was assumed to be 0.07 with an uncertainty of 0.02 \citep[][]{Spoto2015}. The value of $\alpha$ in $\sim$5,350 Monte Carlo trials is on average $\sim$1.01 with a RMS uncertainty of 0.04 as seen in Fig.~\ref{fig.VeritasMC}. The value of $\mu_{\alpha} \; \simeq \; $ 1.01 and a smaller value of $C\; = \; 1.22 \; \times 10^{-5}$ au compared to $C_{EV} \; = \; 2.1 \; \times 10^{-5}$ au calculated from Eq.~\ref{eqn.VEVvsCalphaFinal} assuming $V_{EV}$ $\simeq$ 20 $\mps$, the escape velocity from a parent body with a parent body diameter, $D_{pb} \; = \; 124$ km \citep[][]{Nesvorny2015a} and $\rho \; = \; 1.4 \; \gpcmc$ typical for C-type asteroids \citep[][]{Carry2012} suggests that the spread of fragments in the Veritas family in $a$ vs. $D_r$ space is due almost entirely to the ejection velocity of the fragments. The $D_{pb}$ we use may be wrong because it includes asteroid Veritas as a fragment with a diameter of $\sim$100 km because Veritas may be an interloper in its own family as indicated by impact modedlling \citep[][]{Michel2011}.
 
% BIBLIOGRAPHY -----------------------------------------------------------

\bibliographystyle{icarus}
\bibliography{references}

\end{document}